\documentclass[prb,aps,twocolumn,showpacs]{revtex4-1}

\usepackage{graphicx}
\usepackage{latexsym}
\usepackage{amsmath}
\usepackage{bm}

\newcommand{\be}{\begin{eqnarray}}
\newcommand{\ee}{\end{eqnarray}}

\begin{document}

\title{Two-component Bose gas trapped by harmonic and annular potentials: \\
Supercurrent, vortex flow and instability of superfluidity by Rabi coupling}

\date{\today}

\author{Hayato Ino} 
\author{Yoshihito Kuno} 
\author{Ikuo Ichinose}
\affiliation{%
Department of Applied Physics, Nagoya Institute of Technology,
Nagoya, 466-8555 Japan}

\begin{abstract}
In this paper, we study a system of two-component Bose gas in an artificial
magnetic field trapped by concentric harmonic and annular potentials,
respectively.
The system is realized by gases with two-internal states like the hyperfine
states of $^{87}$Rb.
We are interested in effects of a Rabi oscillation between them.
Two-component Bose Hubbard model is introduced to describe the system, 
and Gross-Pitaevskii equations are used to study the system.
We first study the Bose gas system in the annular trap by varying the 
width of the annulus and strength of the magnetic field, in particular, 
we focus on the phase slip and superflow.
Then we consider the coupled Bose gas system in a magnetic field.
In a strong magnetic field, vortices form a Abrikosov triangular lattice
in both Bose-Einstein condensates (BECs), and locations of vortices 
in the BECs correlate with each other by the Rabi coupling.
However, as the strength of the Rabi coupling is increased, 
vortices start to vibrate around their equilibrium locations.
As the strength is increased further, vortices in the harmonic trap start
to move along the boundaries of the annulus.
Finally for a large Rabi coupling, the BECs are destroyed.
Based on our findings about the BEC in the annular trap, we discuss the origin
of above mentioned phenomena.
\end{abstract}


\maketitle
\section{Introduction}

In the past decade, ultracold atomic systems are one of the most actively
studied fields in physics.
It is expected that these systems can be a quantum simulator for 
various models in quantum physics\cite{qsimu}.
Another aim to study ultracold atoms is to search for new quantum
phenomena whose existence is theoretically predicted.
One of these example is the supersolid that has both a diagonal
solid order and an off-diagonal superfluid order\cite{SS}.

In this paper, we shall consider a system of two-component Bose gas in an 
artificial magnetic field\cite{gauge}.
Each component of the boson corresponds to an internal state of
single boson.
We consider a two-dimensional (2D) system, and 
one component of the boson, which we call $A$-atom, is trapped by a
harmonic potential, and the other boson, called $B$-atom, by an annular trap. 
The atoms are coherently transferred by a Rabi coupling between two
internal states.
It should be remarked that recently closely related atomic system was
realized in experiments\cite{exp1}.
As we show in the present study, interplay of the harmonic
and annular traps generates intriguing phenomena.
That is, 
the system exhibits interesting behavior depending on the strength of the magnetic field, 
inter and intra-repulsions, and the Rabi coupling.
The present study was originally motivated by the findings of co-existing two kinds
of superconductivity in the Sn/Si core-shell clusters\cite{Sn/Si}, i.e.,
we expected that a phenomenon similar to the double superconductivity of 
the Sn/Si core-shell cluster can be realized in a ultracold atomic system.
However in due course, we have realized that the ultracold atomic system 
has strong versatility and controllability.
Much more interesting phenomena are to be observed in that system
than we first expected.
 
This paper is organized as follows.
In Sec.II, we explain the system of the two-component Bose gas trapped 
by harmonic and annular potentials,
and introduce Gross-Pitaevskii equations (GPEs) to study the system.
The bosons are coupled with the artificial magnetic field that is generated, e.g., 
by rotating the system.
The Rabi coupling that induces coherent transference between the $A$ and $B$-atoms
is also introduced.
Rather detailed description of the numerical methods is also given.
In Sec.III, we consider behavior of a single-component BEC in the harmonic and annular
potentials separately.
In particular, we study the ring-shaped BEC in the annular trap in detail.
Behavior of the BEC is strongly depends on width of the annulus and 
strength of the artificial magnetic field.
In the case of the quasi-1D case, a phenomenon similar to the Aharonov-Bohm (AB)
effect takes place as recently observed by the experiments\cite{exp2,expWL}.
On the other hand for a relatively wide ring-shaped BEC, superflows 
are generated in regions near the boundaries.
The superflows are similar to the surface supercurrents in superconductors
in a strong magnetic field.
In Sec.IV, we study the two-component Bose gas in the harmonic and annular
traps with the repulsions and the Rabi coupling.
We fix the intra and inter-repulsions, and vary the strength of the Rabi coupling
to observe how the BECs change. 
For a finite but small Rabi coupling, vortices start to vibrate around their
equilibrium locations.
For a moderate Rabi coupling, vortices of $A$-condensate flow along
the boundaries of the $B$-atom ring-shaped condensate.
As the Rabi coupling is increased furthermore, the BECs are destroyed.
Section V is devoted for conclusion.

\section{BEC$\mbox{s}$ in harmonic and annular traps}
\setcounter{equation}{0}

\subsection{Potential traps and GPE$\mbox{s}$}

\begin{figure}[ht]
\begin{center}
\includegraphics[width=1.8cm]{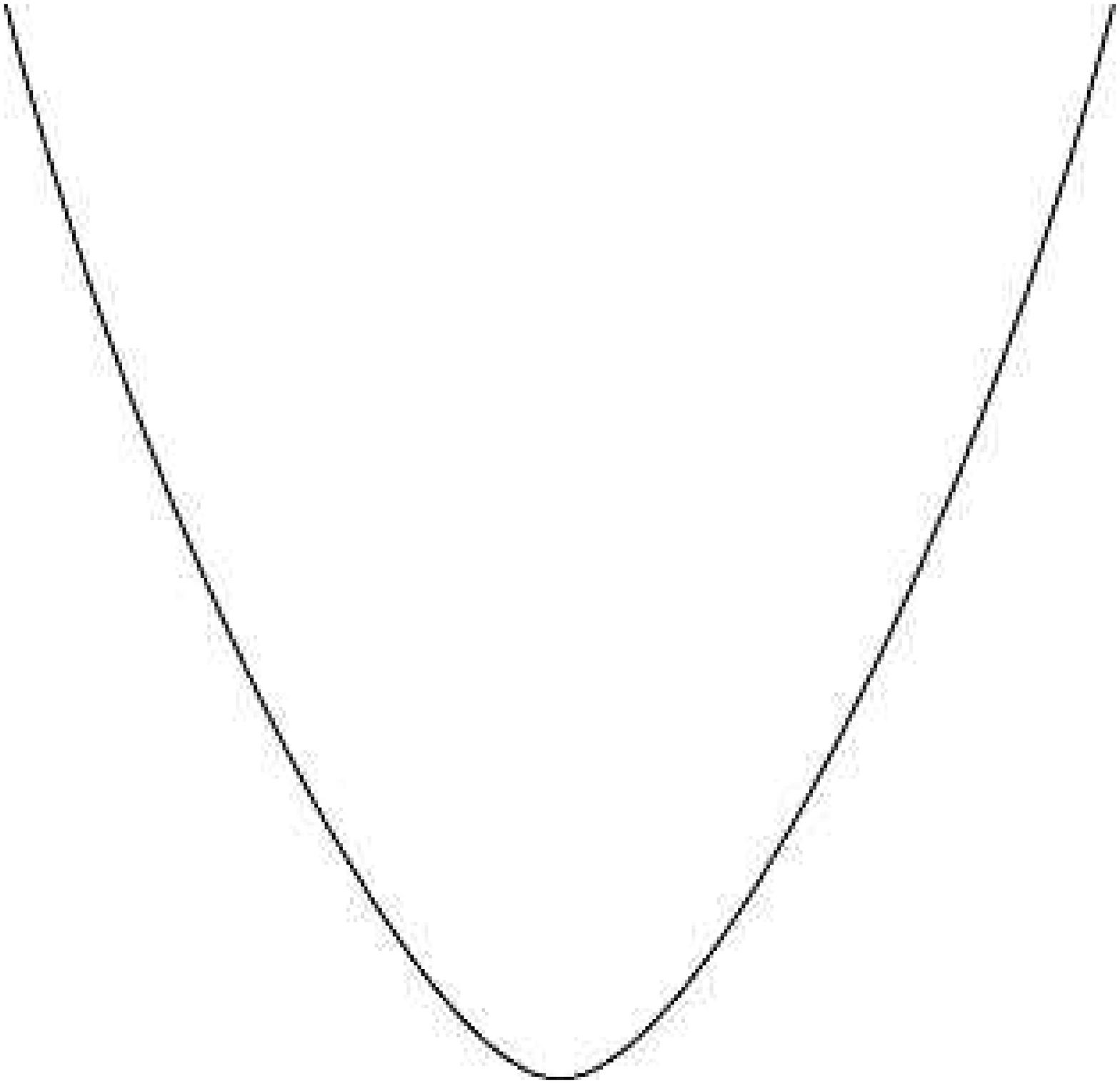} 
\includegraphics[width=2.2cm]{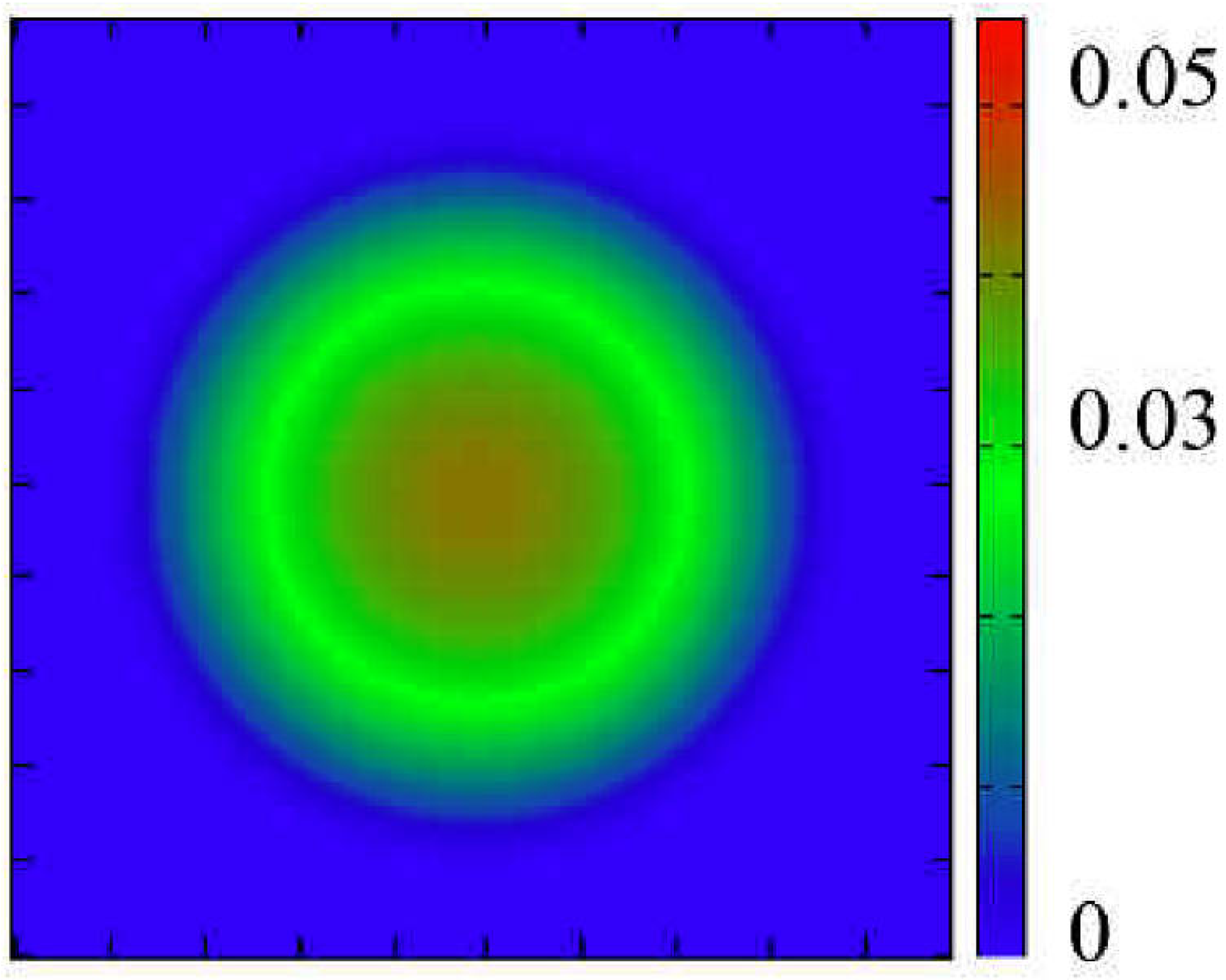}
\includegraphics[width=2cm]{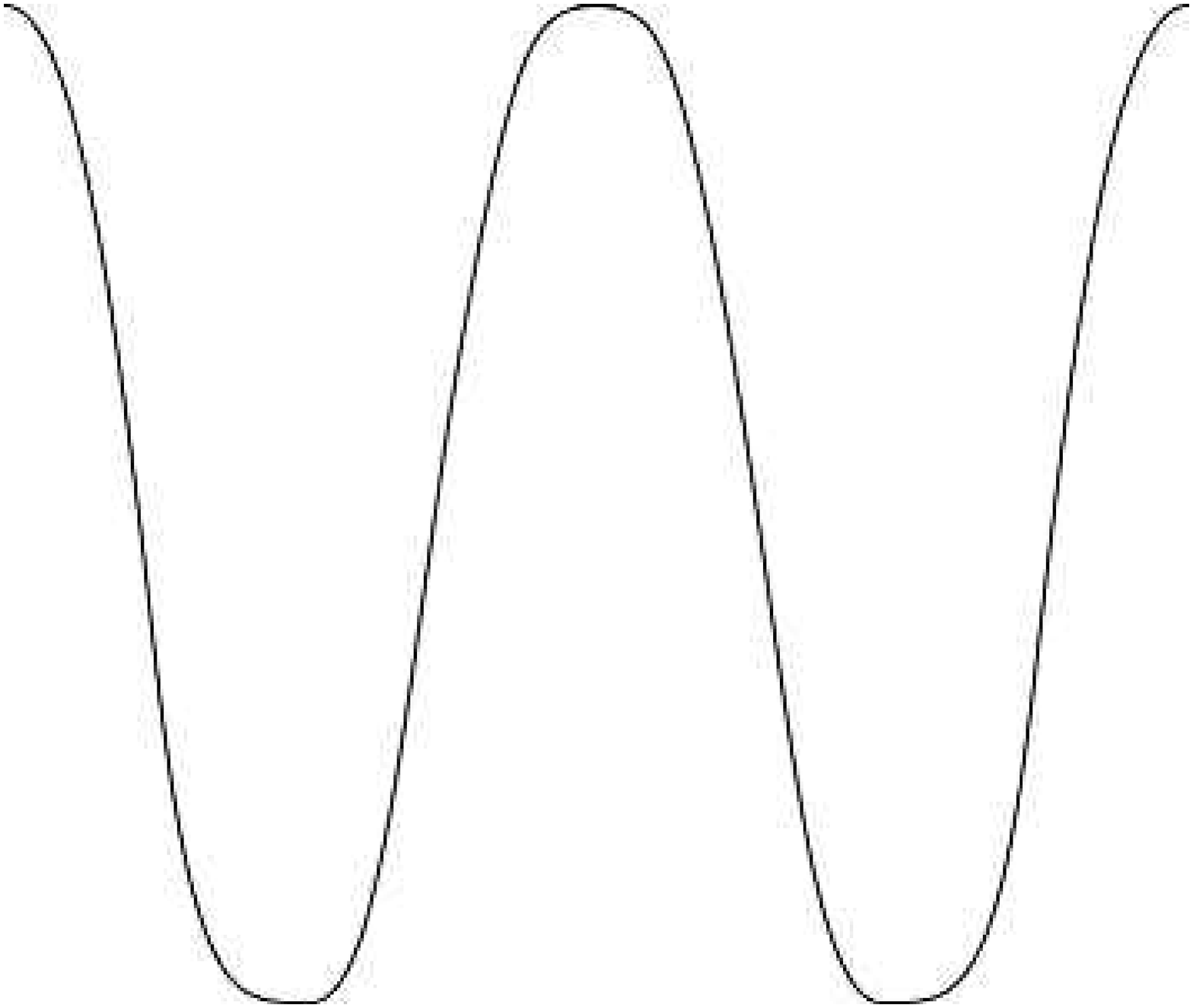}  \vspace{0.5cm}
\includegraphics[width=2.2cm]{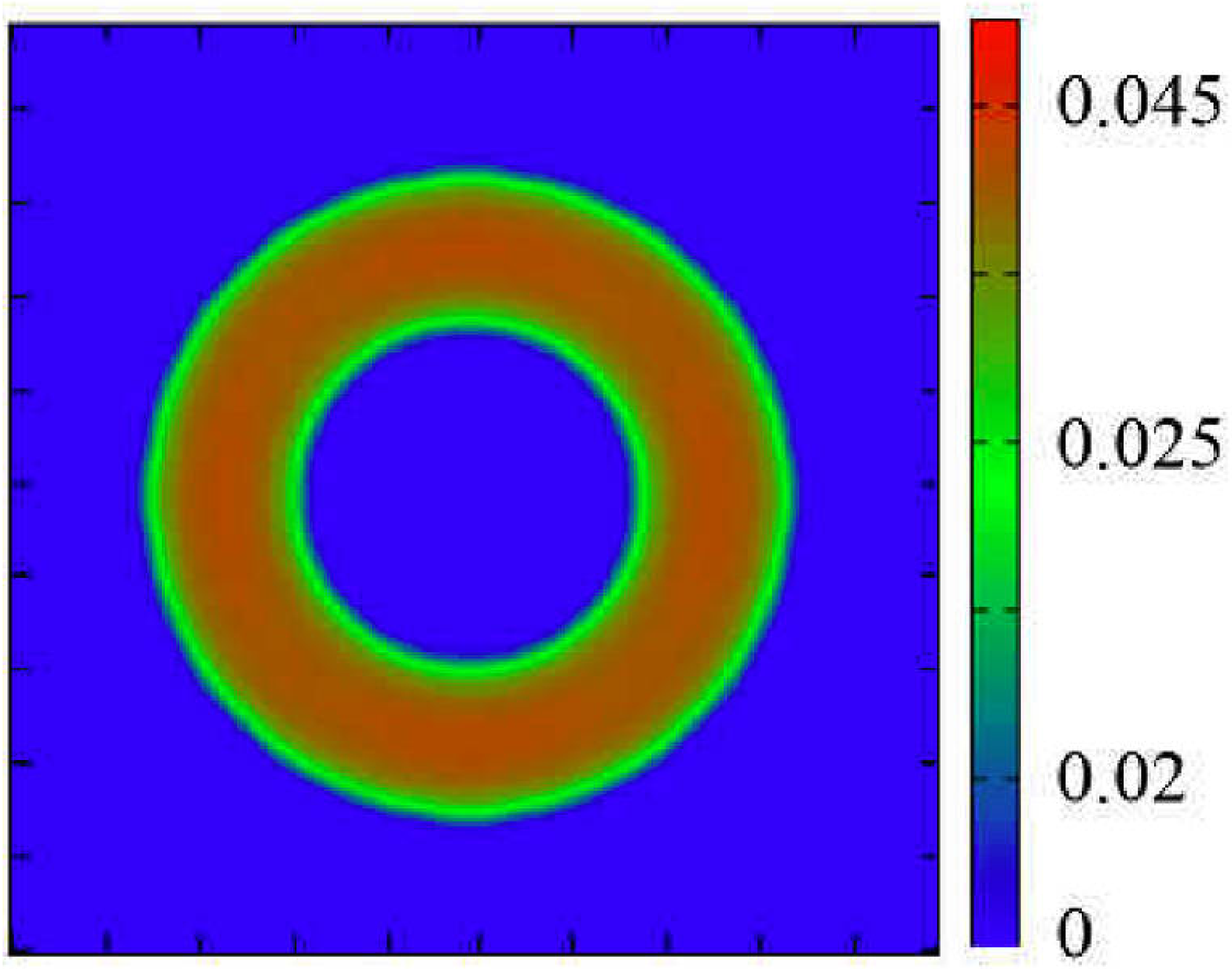} 
\includegraphics[width=3.5cm]{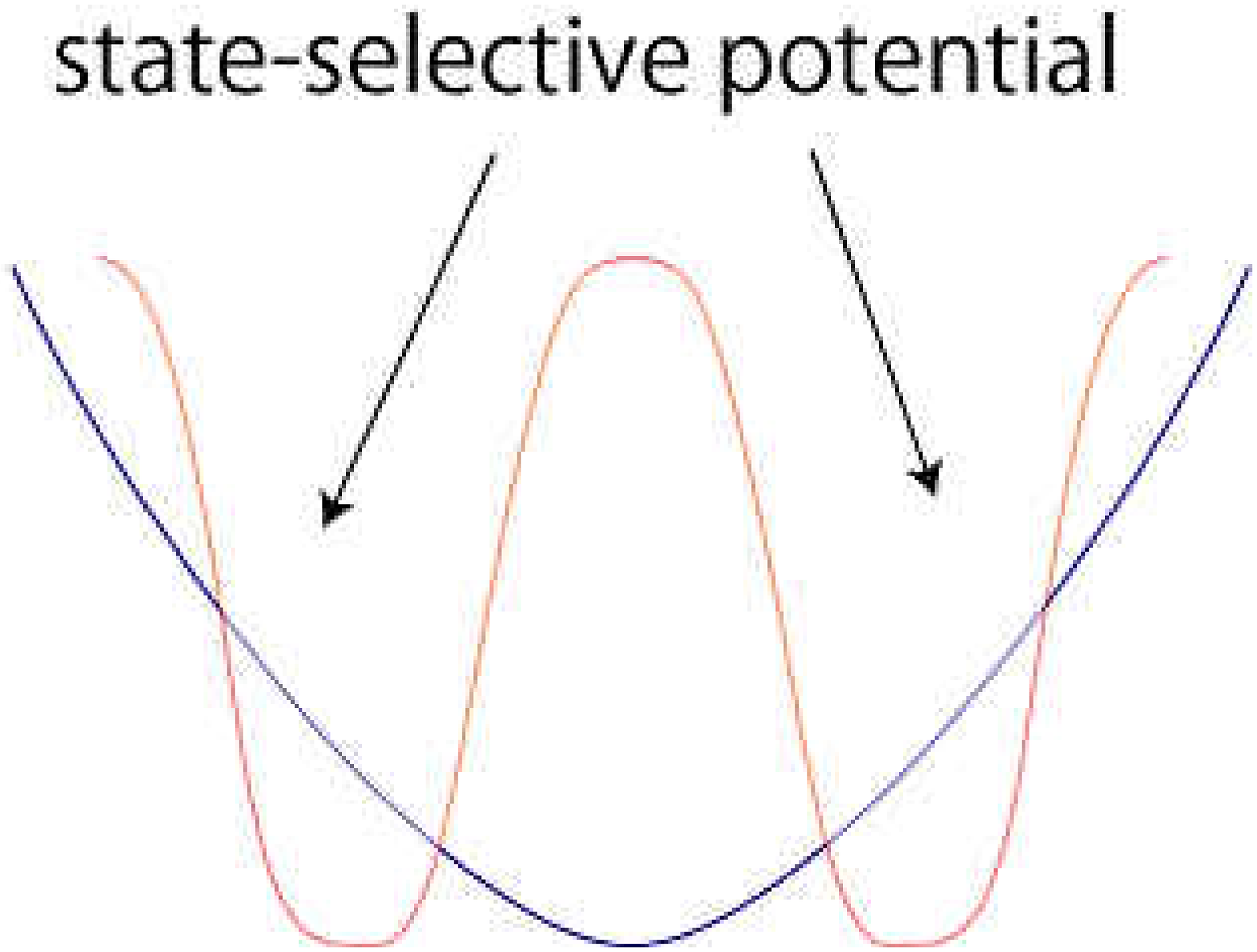} 
\caption{(Color online)
Harmonic and annular traps for two-component Bose gases.
Potentials $V_{A(B)}(x)$ are shown in the upper panel.
The lower panel shows the superposed state-selective potential.}
\label{pot}
\end{center} 
\end{figure}

In this section, we shall derive GPEs describing two-component BECs
trapped by concentric harmonic and annular potentials, respectively.
The two-component bosons mean two-internal states of a single boson like 
the hyperfine states of $^{87}$Rb and
they can be trapped by different potentials by means of the difference in
magnetic/electric properties of them\cite{exp1}.
We call these bosons $A$-atom and $B$-atom, respectively, and denote
the state-selective potential by $V_{A(B)}(x)$, where $x=(x_1,x_2)$ is 
the two-dimensional (2D) coordinate.
See Fig.\ref{pot}.

GPEs of condensate-field
$\psi_A(x)$ and $\psi_B(x)$ for the coupled $A$ and $B$-atoms system 
are given by
\begin{eqnarray}
(i-\gamma)\hbar {d\psi_A \over dt}  
&=\Big(-{\hbar^2 \over 2m}\sum_{i=1,2}(\partial_i-iA_i)^2-\mu   \nonumber \\
&\hspace{-1cm}+R\bar{\psi}_A\psi_A 
+U\bar{\psi}_B\psi_B+V_A\Big)\psi_A +g\bar{\psi}_B,
\nonumber \\
(i-\gamma)\hbar {d\psi_B \over dt}    
&=\Big(-{\hbar^2 \over 2m}\sum_{i=1,2}(\partial_i-iA_i)^2-\mu  \nonumber \\
&\hspace{-1cm}+R\bar{\psi}_B\psi_B  
+U\bar{\psi}_A\psi_A+V_B\Big)\psi_B+g\bar{\psi}_A,
\label{GPEq}
\end{eqnarray}
where $R(U)$ is the intra(inter)-species repulsion and $g$ is the 
Rabi coupling\cite{harmonic}.
$A_i$ is a vector potential for a uniform artificial magnetic field, which is
created by rotating the system or using lasers.
The parameter $\mu$ is the chemical potential, which controls the particle number, 
and $\gamma$ is a dissipative parameter.
We put $\gamma=0.01$ for most of the practical calculation.
To solve the GPEs (\ref{GPEq}), we employ the Crank-Nicolson
method, and use the symmetric gauge for the vector potential $A_i$, i.e.,
$(A_1(x),A_2(x))=(\pi f x_2, -\pi f x_1)$.
Therefore the magnetic flux per unit area is $2\pi f$. 
For a rotating system,
$f\propto m\Omega$, where $\Omega$ is an angular velocity of the rotation.
When both the Rabi coupling $g$ and inter-species repulsion $U$ are
vanishing, the GPEs (\ref{GPEq}) are two independent equations, and
the dynamics of the solutions are governed simply by the potentials $V_A(x)$,
$V_B(x)$ and the strength of the intra-repulsion $R$.
As we show in the following sections, the inter-coupling $U$ and
the Rabi coupling $g$ strongly influence the structure of the groundstate and 
low-energy excitations of the coupled system.
In particular, interplay of the Rabi coupling and the potentials $V_{A(B)}$ plays a
very important role.

We briefly explain the numerical methods to solve
the GPEs (\ref{GPEq}).
As the unit of the length, we used the spatial cutoff $a=0.5\mu$m.
Every length in this paper is measured in this unit.
For the energy unit, we used Na mass, $m_N=4.0\times 10^{-27}$Kg,
and then ${\hbar^2 \over m_Na^2}=80$nK.
Every energy is measured in this unit.
Typical time slice used for solving the GPEs (\ref{GPEq}) is $\tau=10^{-6}$sec.
The dissipative damping factor $\gamma=0.01$, as explained above,
and we used the Crank-Nicolson method to evaluate $\psi_{A(B)}(t+\tau)$
from $\psi_{A(B)}(t)$.
The total numbers of particles were assumed $10^6\sim 10^5$ 
as in the typical experiments.
In the initial state of the time evolution, the ratio of the total number of atoms in the
harmonic and annular traps, $N_{\rm h}$ and $N_{\rm a}$, was set 
$N_{\rm h}/N_{\rm a}=1$.
After each time evolution, we normalized the wave functions $\psi_{A(B)}(t)$
to keep the total number of atoms constant.
Typical number of the time steps is $\sim 10^6$, and therefore typical time scale
of numerical simulations is $\sim 1$sec, which is comparable with the experiments.

\section{Numerical results of GPE$\mbox{\rm\bf s}$ for single-component
BEC in magnetic field}
\setcounter{equation}{0}

\subsection{Single-component BEC in artificial magnetic field}
\begin{figure}[ht]
\begin{center}
\includegraphics[width=8cm]{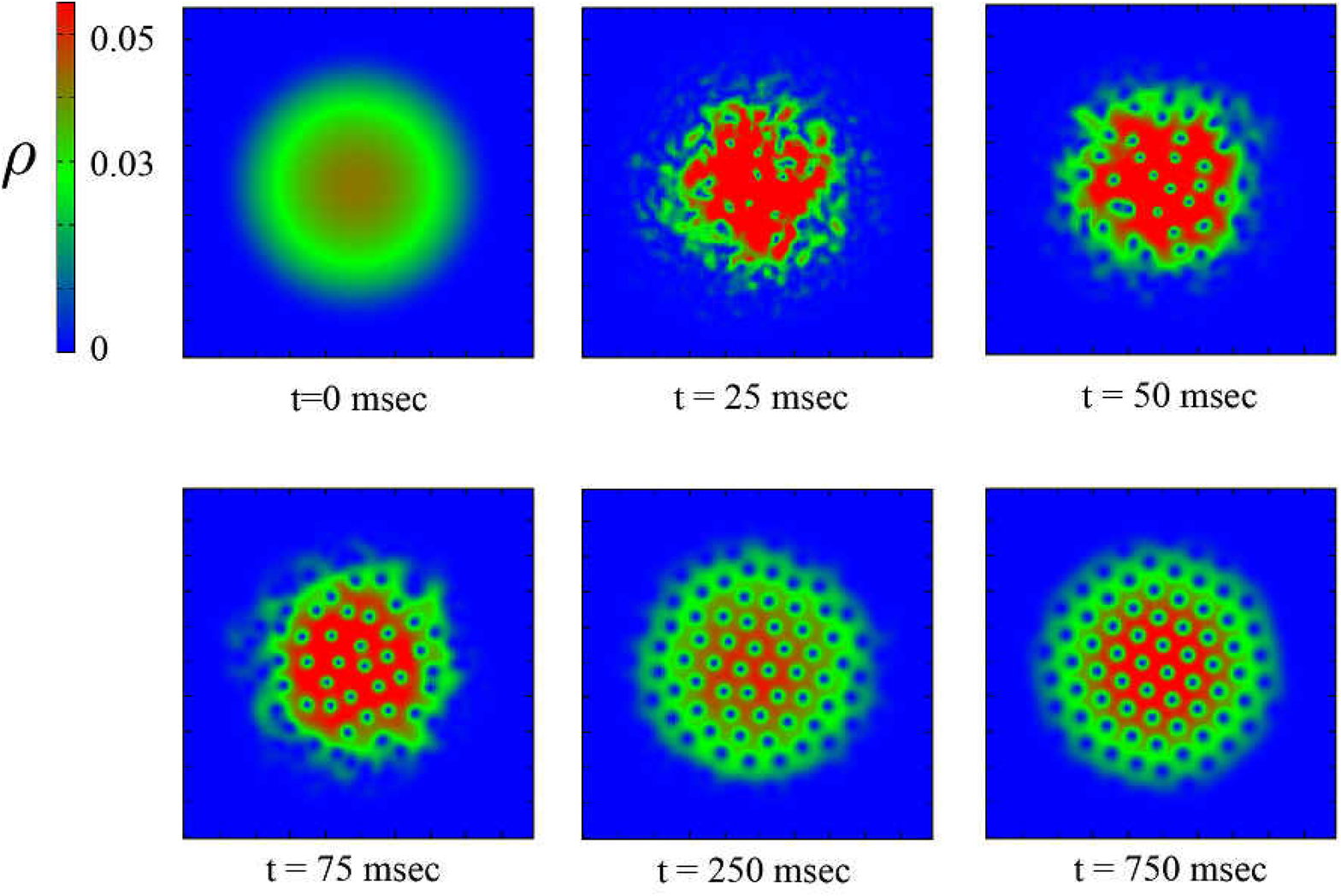} \vspace{0.5cm} \\
\includegraphics[width=8cm]{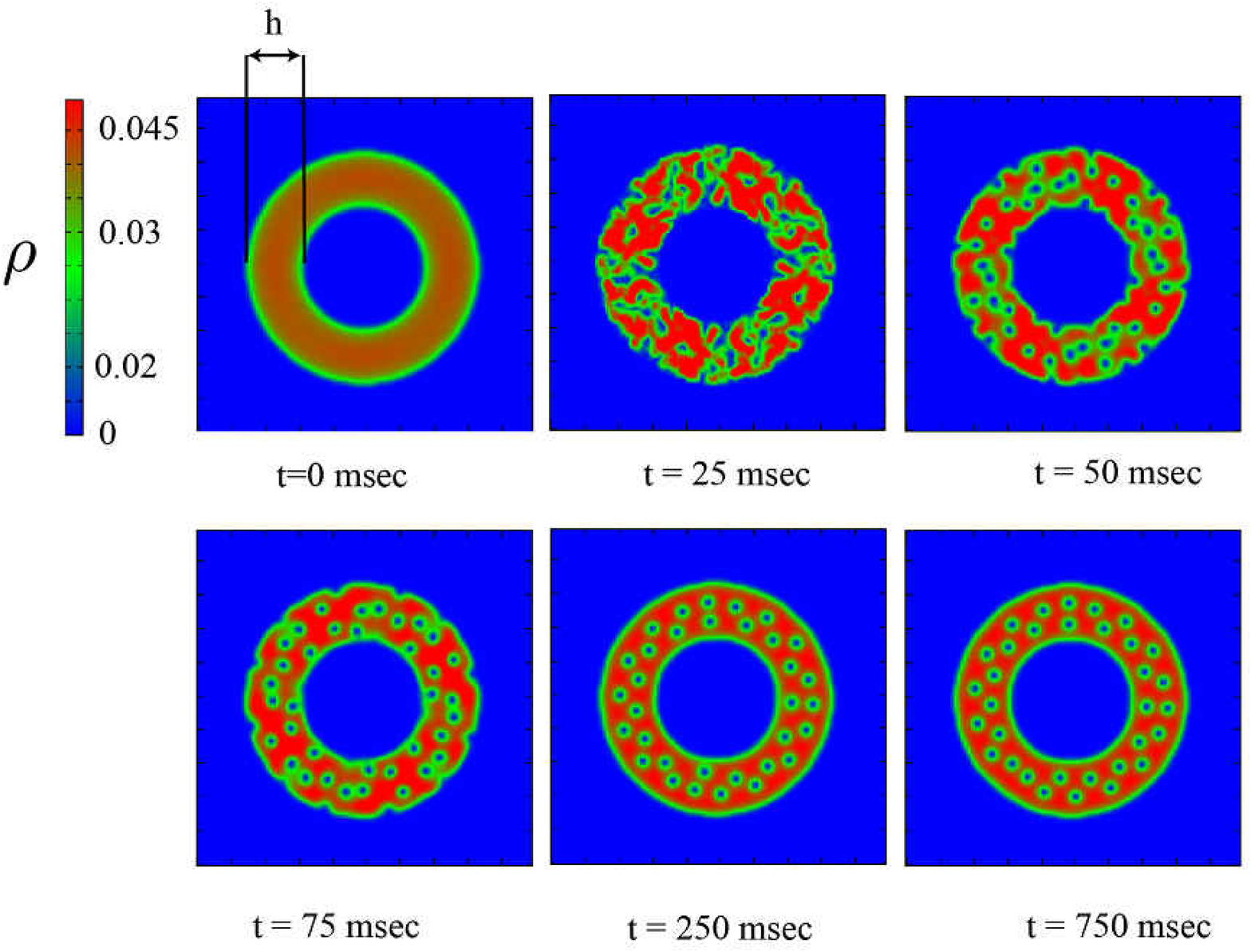} \vspace{0.5cm} \\
\includegraphics[width=8cm]{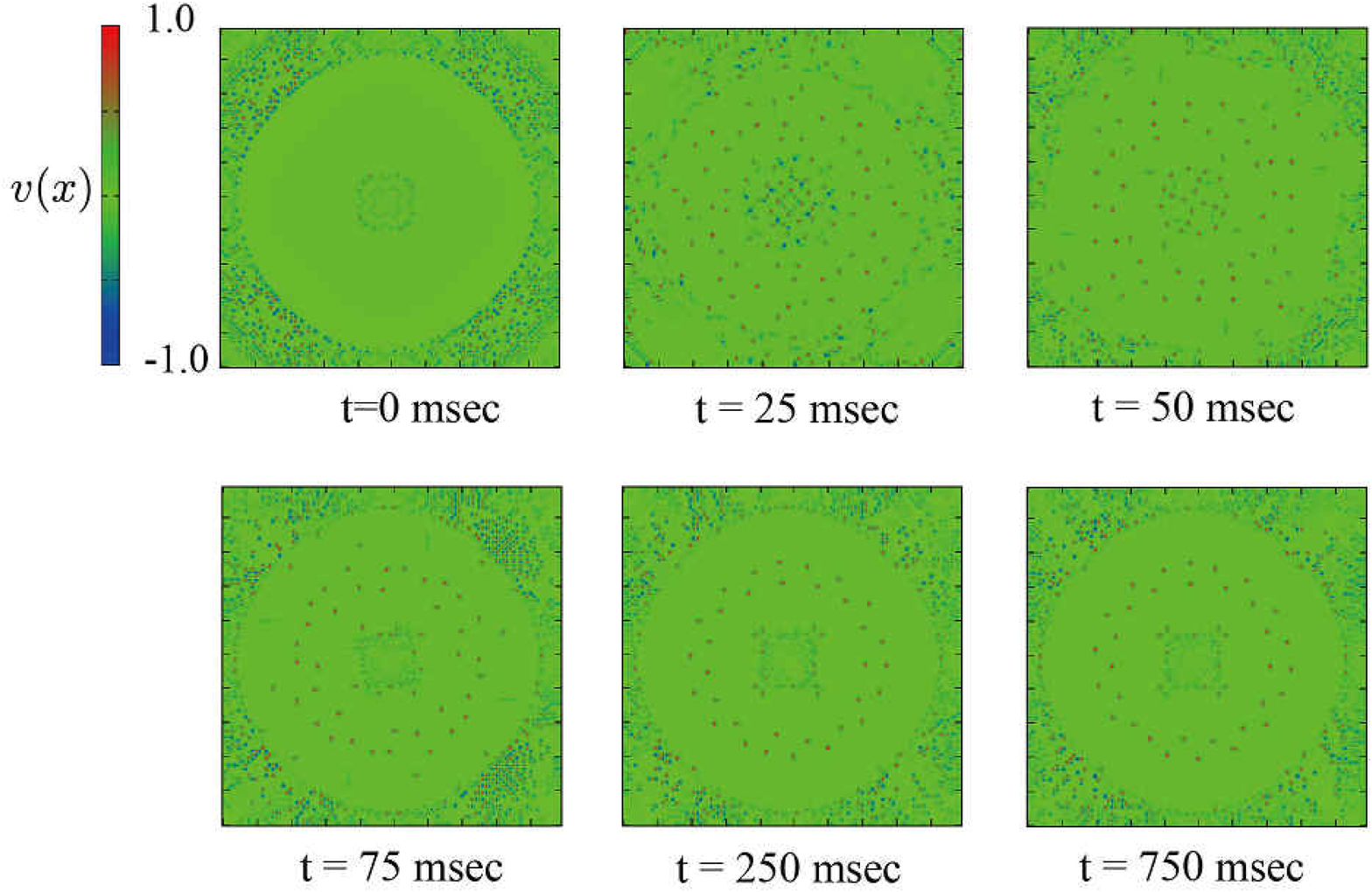} 
\caption{(Color online)
The case of the vanishing Rabi coupling and inter-repulsion.
Snapshots of density of the BECs in the harmonic and annular traps,
and the locations of vortices $v(x)$ in the annular BEC.
Abrikosov vortex lattice forms in each BEC. 
Parameters are $h=20$, ${\hbar^2 \over m}=1$ and $f=0.02$.}
\label{snap1}
\end{center} 
\end{figure}

In this section, we show the numerical results for the
GPEs (\ref{GPEq}) of the BECs in both the harmonic and annulus traps.
We set ${\hbar^2 \over m}=1$.
For the case of $U=g=0$, which corresponds to the two independent BECs trapped by
each potential, a stable vortex lattice is expected to form in each BEC for a finite 
intra-repulsion $R>0$ and a strong magnetic field.
We consider the case with the external magnetic field $f=0.02$, which is
relatively strong, and $R=1$.
In Fig.\ref{snap1}, snapshots are shown for the BEC confined in the harmonic potential.
Fig.\ref{snap1} also shows the time evolution of the BEC confined in the annular trap
of the inner radius $\ell=20$ and the width $h=20$, which is {\em larger} than 
the healing length of the BEC.
To measure vorticity $v(x)$ of the BEC with an order parameter
$\psi=\sqrt{\rho}e^{i\theta}$, we use the following definition,
\begin{equation}
v(x)={1 \over 2\pi}\mbox{rot} \; \theta(x).
\label{vor}
\end{equation}

It is obvious that an Abrikosov lattice forms even for the BEC in the
annular trap as verified by both the density and vorticity snapshots
in Fig.\ref{snap1}.
We have verified that 
as the strength of the magnetic field is increased, the density of vortex
increases in both the harmonic and annular traps.
The snapshots in Fig.\ref{snap1} clearly exhibit that BEC first becomes unstable 
by the sudden application of the external magnetic field
and then subsequently vortices appear near the boundaries of the
BEC, enter the bulk of the BEC and finally form an Abrikosov lattice\cite{kasamatsu}.
Formation of the Abrikosov lattice of vortex in the annular BEC implies the existence of 
stable surface superflow as we verified in the following section.

\begin{figure}[ht]
\begin{center}
\includegraphics[width=8cm]{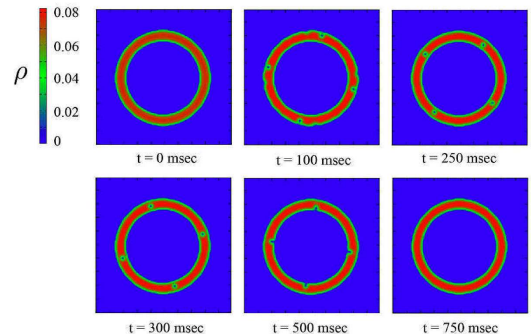} 
\caption{(Color online)
Snapshots of BEC density in the annular trap with $h=10$.
Application of sudden magnetic field generates vortices but they move to
the outside of the BEC.
As a result, Abrikosov lattice does not form.
Parameters are ${\hbar^2 \over m}=1$ and $f=0.02$.}
\label{snap1D}
\end{center} 
\end{figure}

It is interesting to see how vortices behave in a narrower BEC confined
in a quasi-one-dimensional (q1D) annulus.
We studied the BEC in the annular trap with $h=10$, and show the numerical results 
in Fig.\ref{snap1D}. 
Vortices are first generated by the sudden application of the artificial magnetic field
with $f=0.02$,
but they move to the outside of the BEC, and no Abrikosov lattice forms.
We also studied the BEC in the annular trap with $h=5$ and found that no
vortices are generated by the external magnetic field.
In the case with the magnetic field $f=0.02$ and $h<10$, the boson system
can be regarded as a 1D system, and therefore it is expected that
the superflow behaves differently from that in the 2D case with larger $h$.
In the following section, we shall verify this expectation.

\subsection{BEC in annular trap, weak link and phase slip}

\begin{figure}[ht]
\begin{center}
\includegraphics[width=4cm]{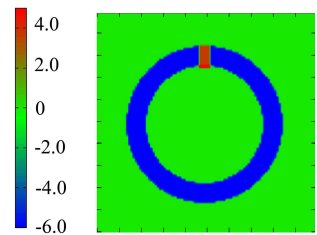} 
\caption{(Color online)
Annular trap with a weak link.
Potential for the weak link is explicitly given by Eq.(\ref{potwl}). 
}
\label{PWL1}
\end{center} 
\end{figure}

The results in the previous section were obtained for the magnetic field $f=0.02$,
for which the corresponding magnetic flux in the central hole region of 
the annulus, $\Phi_{\rm in}$, is about $8\pi$ and very large.
As a result, the vortex lattice forms in the bulk of the ring-shaped condensate 
with the width $h>20$.
On the other hand
in most of the practical experiments on the rotating BEC in an annular trap,
the total magnetic flux $\Phi_{\rm in}$ is about $\pi\sim 4\pi$,
and a phenomenon similar to the AB effect was observed.
In some experiments, a weak link (WL) was produced in condensates by
a focused blue-detuned laser field and a rotating BEC was produced
by stirring the condensate by the WL.
Recently, a specific method was invented and used to measure the
phase of the ring-shaped BEC directly\cite{expWL}.

In order to describe the above mentioned experimental setup, we add a WL
potential $V_{\rm wl}$ in addition to $V_B(x)$.
In the rotating frame, the WL potential is given as 
\begin{equation}
V_{\rm wl}=w\theta(x_1+\delta)\theta(\delta-x_1)
\theta(x_2-\ell)\theta(\ell+h-x_2),
\label{potwl}
\end{equation}
where $w$ is the height of the potential barrier, $\delta$ is its width,  
and we typically take $\delta=5$.
See Fig.\ref{PWL1}.

We expect that as the width of the annulus $h$ is varied, behavior of the
annular condensate changes gradually.
In particular, in the sufficiently small $h$
compared to the magnetic length determined by $f$ and also the healing length
of the BEC, the system is effectively described by a 1D model.
Corresponding to the experiments, we consider the case such that
$\Phi_{\rm in}\sim 2\pi$ (i.e., $f=0.005$).
In this case, it is expected that no vortices exist in the bulk of the condensate
even for a wide annulus like $h\sim 20$.

We first study the ring-shaped condensate with 
$h=5$ and ${\hbar^2 \over m}=10$.
We verified that for $w=0$ and $f=0$ (i.e., a homogeneous condensate 
in the vanishing magnetic field) the condensate has a uniform phase,
as is expected.

\begin{figure}[ht]
\begin{center}
\includegraphics[width=8cm]{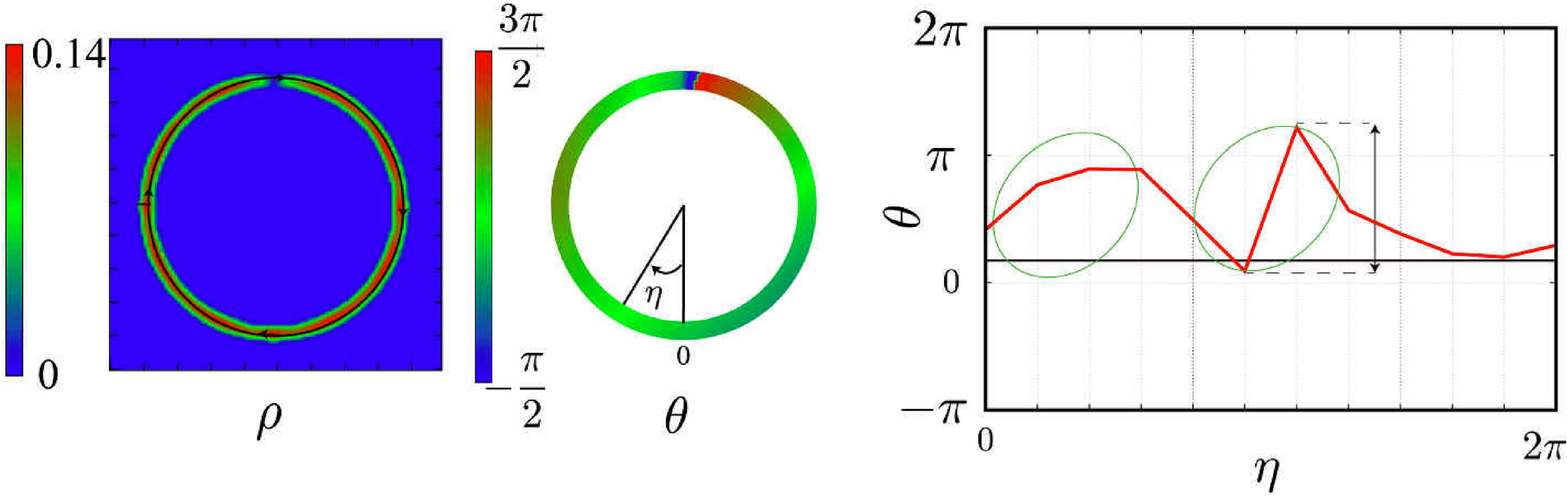} 
\includegraphics[width=8cm]{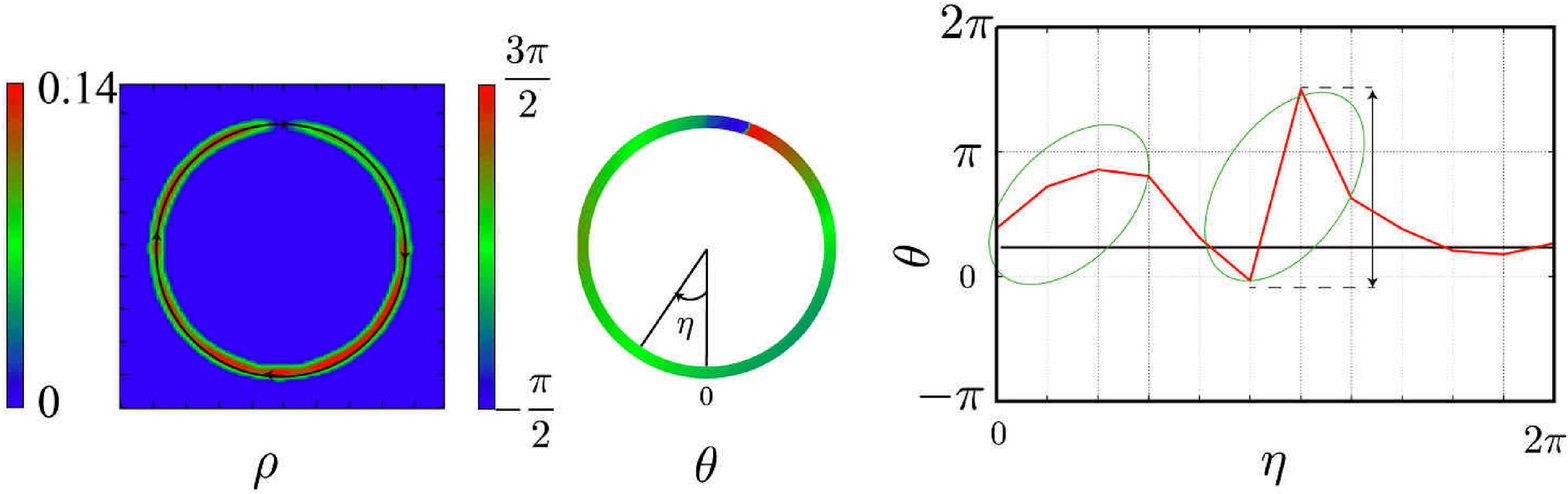} 
\caption{(Color online)
Density $\rho$ and phase $\theta$ of the ring-shaped condensate with WL.
Height of the WL, $w=10$ (upper panel) and $w=100$ (lower panel), $f=0.005$
and ${\hbar^2 \over m}=10$.
The phase of BEC increases in the WL and also the opposite side of the WL.}
\label{DPWL1}
\end{center} 
\end{figure}

For the numerical study,
we first prepared the homogeneous condensate for $h=5, w=0$ and $f=0$, 
and then we suddenly applied the WL of $w=10(100)$ and the magnetic field $f=0.005$.
Solution of the GP equation shows that the condensate was first disturbed by
the applied magnetic field and WL, but became stable by the effect of the dissipation.
The density and phase of the final states are shown in Fig.\ref{DPWL1}.

\begin{figure}[ht]
\begin{center}
\includegraphics[width=6cm]{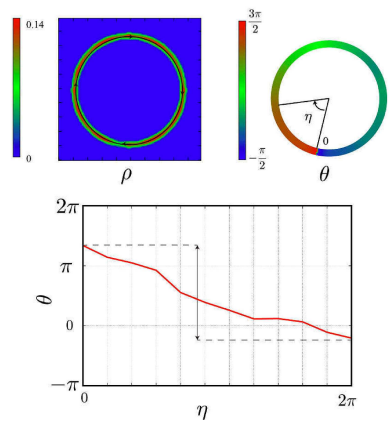} 
\caption{(Color online)
From the states in Fig.\ref{DPWL1}, we decreased $f$ and $w$.
Density and phase of the ring-shaped condensate without 
the magnetic field and WL.
It is observed that the state with the unit winding is realized.
}
\label{DPWL0}
\end{center} 
\end{figure}

The results in Fig.\ref{DPWL1}, in particular the phase of the condensate, should
be compared with the experimental observation.
As reported in Ref.\cite{expWL} for the rotating BEC, the phase of the 
condensate exhibits a sudden increase in the WL.
As the superfluid (SF) velocity $\vec{v}_{\rm s}$ in the {\em rotating frame} is given by 
$\vec{v}_{\rm s}={\hbar \over m}\nabla \theta(x)$, the above behavior
of $\theta(x)$ means that SF in the WL flows in the opposite direction
against to the bulk superflow.
This result is in agreement with the experimental observation in Refs.\cite{expWL},
and theoretical studies in Refs\cite{theor1,theor2}.
However, the calculation in Fig.\ref{DPWL1} also shows the existence of 
smooth increase of the phase in rather wide region of the opposite side of the WL.
This state in Fig.\ref{DPWL1} is the lowest-energy state, and the result indicates that 
a phase increase in the WL induces another phase increase in its opposite
side of the ring-shaped condensate.
The supercurrent $\vec{j}$ {\em observed in the static frame},
$\vec{j}={\hbar \rho \over m}(\nabla -\vec{A})\theta$, on the other hand, exhibits 
no anomalous behavior as we show in the following section.

\begin{figure}[ht]
\begin{center}
\includegraphics[width=8cm]{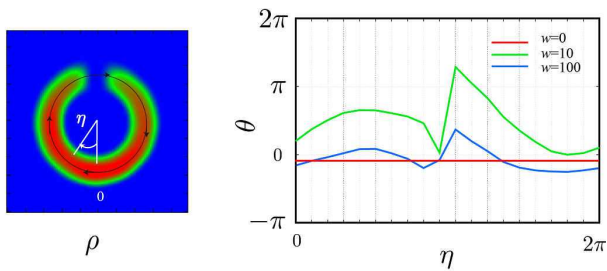} 
\caption{(Color online)
Density and phase of the ring-shaped condensate with WL.
Height of the WL $w=0, 10$ and $100$, $h=20$, $f=0.005$ and ${\hbar^2 \over m}=10$.
}
\label{DPLH}
\end{center} 
\end{figure}

Finally from the state shown in Fig.\ref{DPWL1}, we gradually decreased 
the magnetic field and WL to $f=w=0$, and then observed the density and
phase of the condensate.
The results are shown in Fig.\ref{DPWL0}.
This is obvious that homogeneous superflow is generated with the unit
circulation\cite{expSC}.
We also studied the case of wider annular trap.
The results are shown in Fig.\ref{DPLH}.
For large $w=100$, the phase of the condensate is almost constant except in the
WL region.
This result means that the BEC moves with the WL, as is expected.

\begin{figure}[ht]
\begin{center}
\includegraphics[width=8cm]{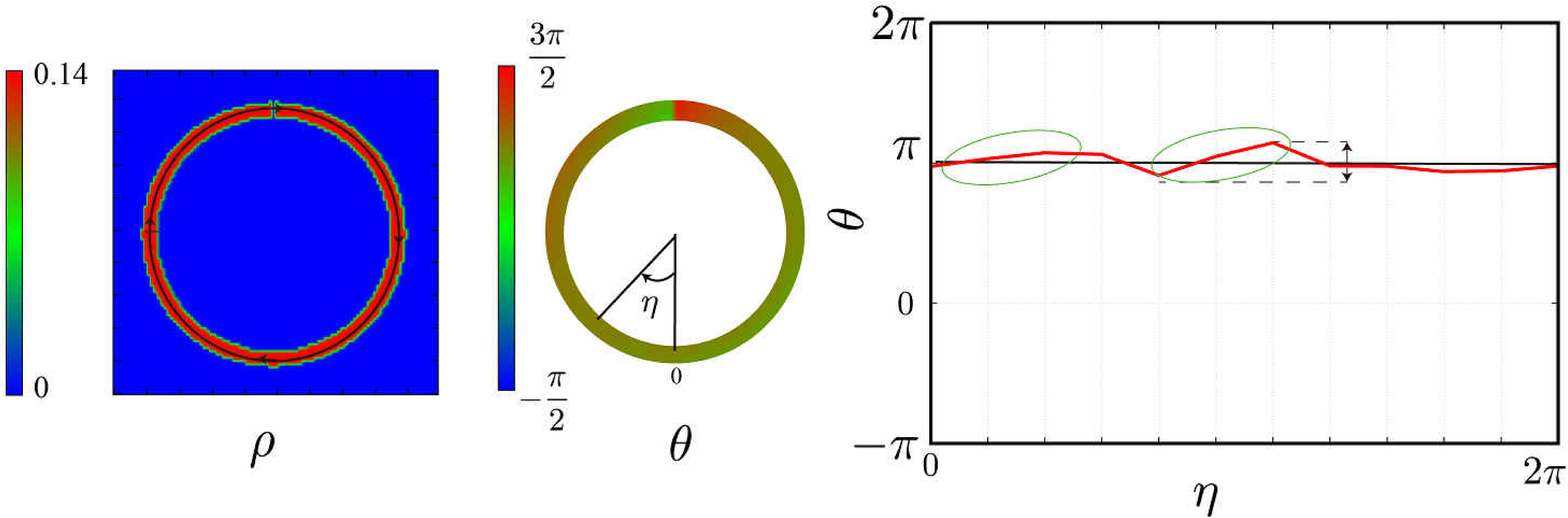} 
\includegraphics[width=8cm]{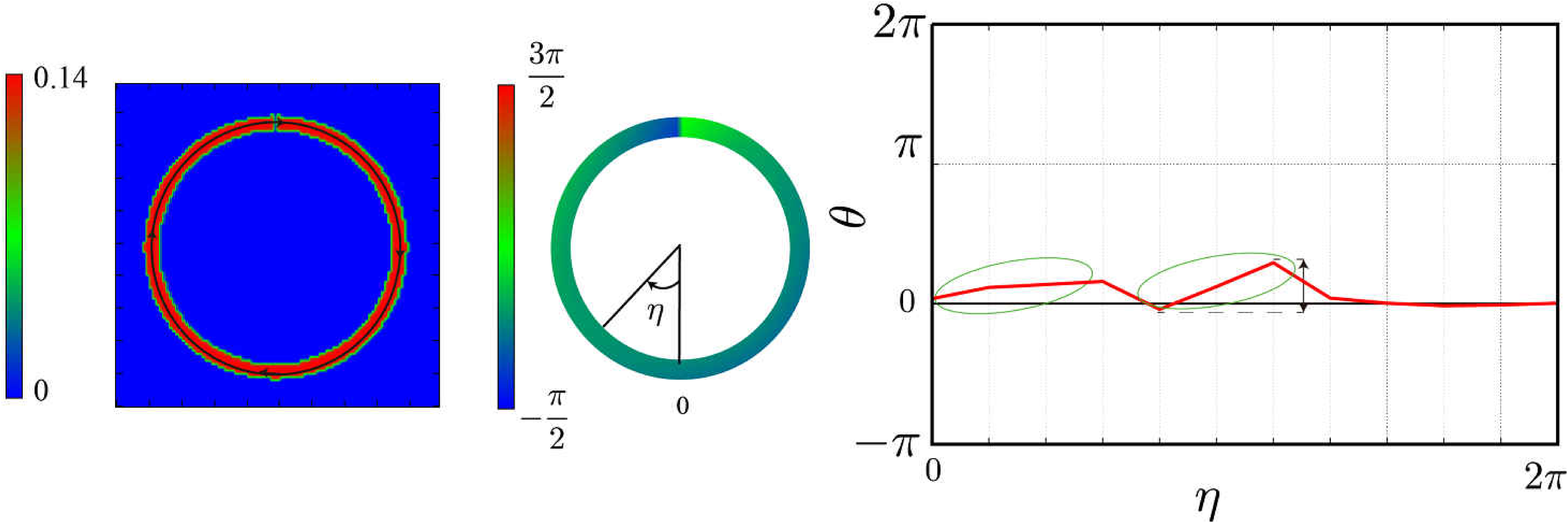} 
\caption{(Color online)
Density and phase of the ring-shaped condensate with the WL.
Height of the WL $w=10$ and $100$, $f=0.005$ and ${\hbar^2 \over m}=0.1$.
}
\label{DPWL2}
\end{center} 
\end{figure}

For a small hopping case like ${\hbar^2 \over m}=0.1$, the obtained results
are shown in Fig.\ref{DPWL2}.
It is obvious that the deviation of the phase is small compared to the case with
${\hbar^2 \over m}=10$.
For small hopping amplitude, the condensate moves with almost the same velocity
with that of the WL, i.e., the SF velocity in the rotating frame is
very small.

\subsection{Supercurrents}

In this section, we shall study the supercurrent $\vec{j}(x)$
in the ring-shaped condensate.
In the previous sections, we showed that there are two distinct
states of the BEC in an external magnetic field, i.e.,
in one case vortices exist in bulk of the condensate because of a
relatively large width $h$ and also a strong magnetic field.  
The other case corresponds to the thiner condensate and/or
condensate in a weak magnetic field in which no bulk vortices exist.
Different behavior of the supercurrent is expected for the above two cases.

\begin{figure}[ht]
\begin{center}
\includegraphics[width=8.5cm]{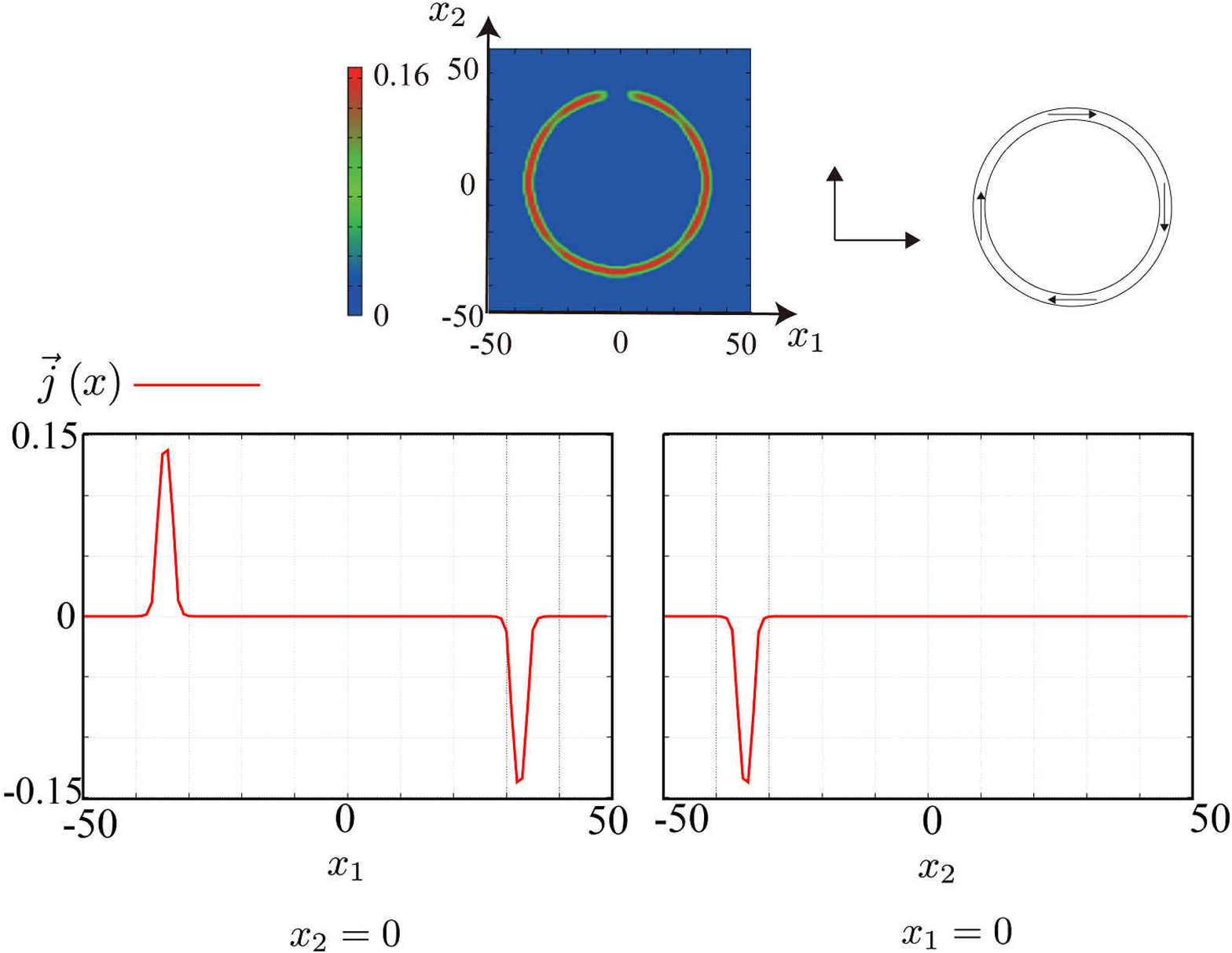} 
\caption{(Color online)
Supercurrent of the ring-shaped condensation in the static frame.
Parameters are ${\hbar^2 \over m}=10$, $f=0.02$, $h=5$ and $w=10$.
}  \vspace{-0.5cm}
\label{PC1}
\end{center} 
\end{figure}
\begin{figure}[ht]
\begin{center}
\includegraphics[width=8.5cm]{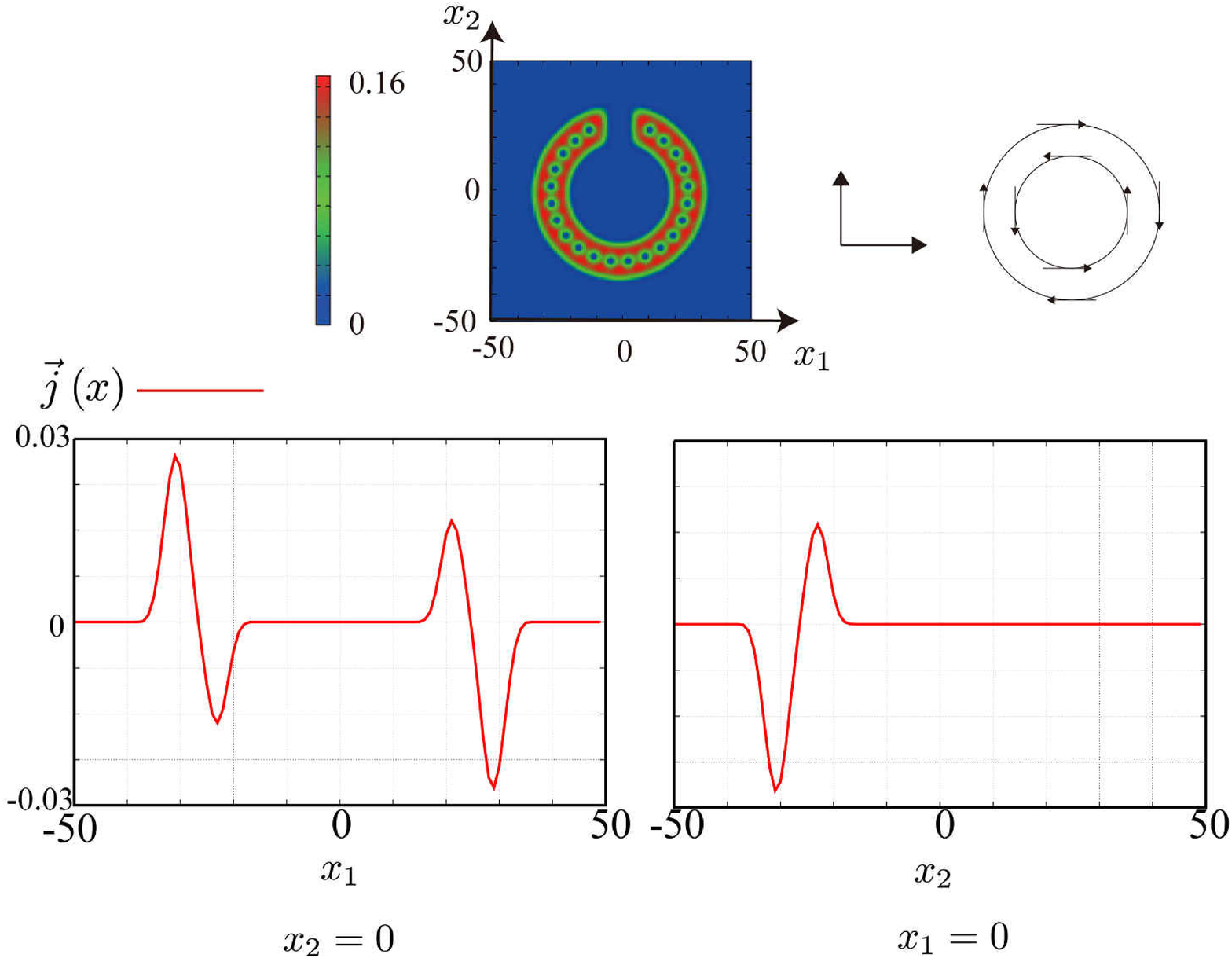} 
\caption{(Color online)
Supercurrent of the ring-shaped condensation in the static frame.
Parameters are ${\hbar^2 \over m}=10$, $f=0.02$, $h=20$ and $w=10$.
}   
\label{PC2}
\end{center} 
\end{figure}

We first consider the case of the ring-shaped condensate with $f=0.02$, $h=5$
and the WL $w=10$.
The result is exhibited in Fig.\ref{PC1}, which shows the supercurrent
$\vec{j}$ in the azimuthal direction.
It is obvious that the supercurrent flows clockwise.
Near the WL, the current is vanishingly small because of the 
low density of the condensate.
As the supercurrent is conserved in the state of a static density, 
there exists supercurrent in the radial direction near the WL
as the previous study showed\cite{theor2}.

On the other hand for the BEC in which vortices exist in the bulk,
the supercurrent flows as shown in Fig.\ref{PC2}.
From the result, it is obvious that the there exist surface supercurrents
near the inner and outer boundaries and they are stabilized by the
vortex lattice in the bulk.
This behavior resembles the ordinary second-kind superconductor in a strong external
magnetic field.
As in the quasi-one-dimensional case in Fig.\ref{PC1}, the supercurrent has to
have a radial component near the WL.
This surface supercurrent plays an important role when
two BECs in the harmonic and annular traps interact with each other through the
Rabi coupling, as we show in Sec.IV.

\subsection{Inter-species repulsion vs. Rabi coupling}

\begin{figure}[ht]
\begin{center}
\includegraphics[width=9cm]{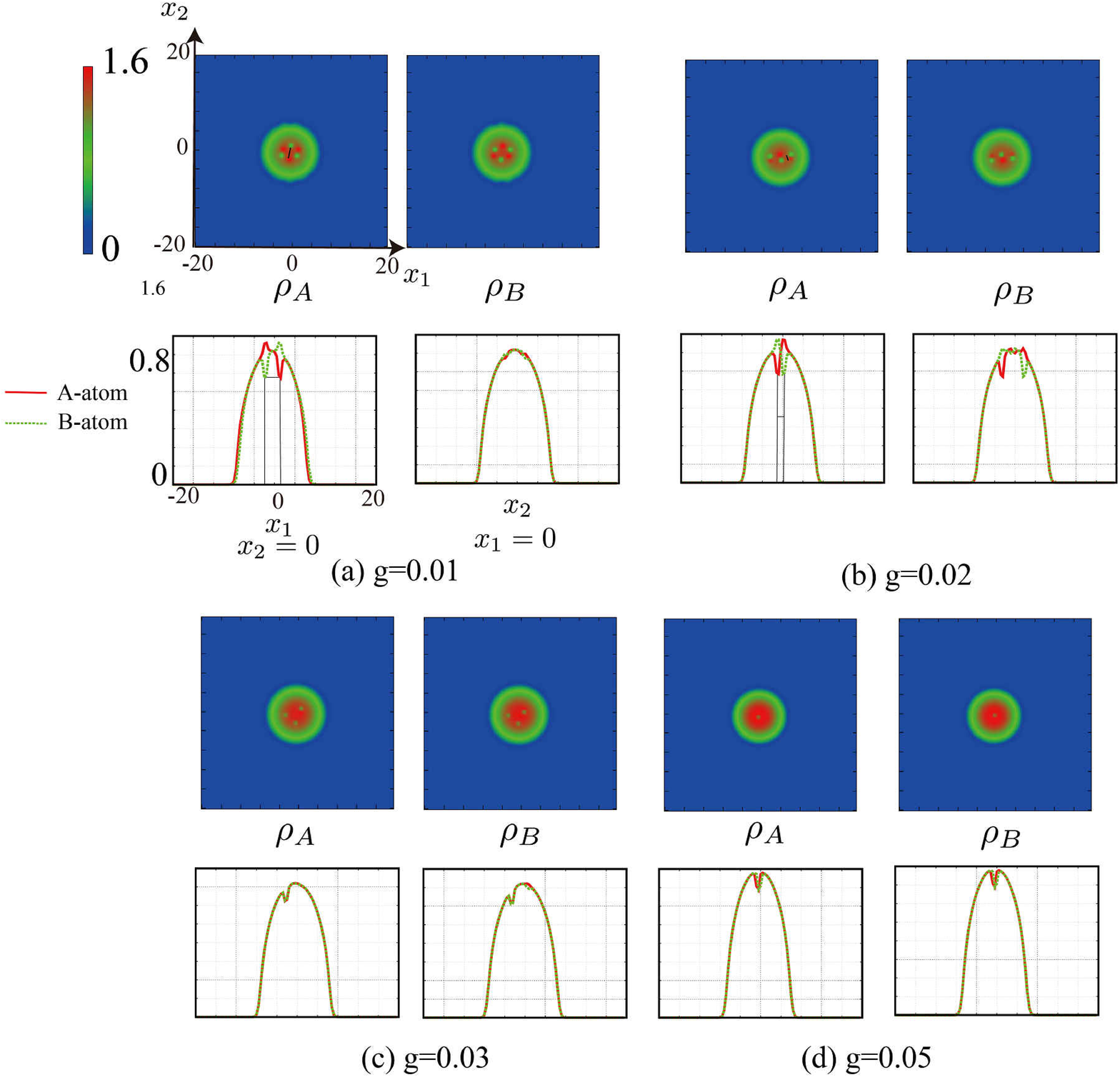} 
\caption{(Color online)
Density overlap of $A$ and $B$-atoms' condensates.
Each BEC contains three vortices.
As the Rabi coupling is increased, the larger overlap appears.
The inter-species repulsion $U=1.1$.
}
\label{repRabi}
\end{center} 
\end{figure}

By the practical calculation, we verified that the Rabi coupling tends to
make the $A$ and $B$-atom condensates overlap with each other.
On the other hand, the inter-species repulsion makes two condensates immiscible.
In this section, we study how the overlap of two condensates varies as the ratio of the
Rabi coupling and the inter-species repulsion, ${g\over U}$ changes.
To this end, we trap both the $A$ and $B$-atoms in the same harmonic
type potential.
The depth of the potential is controlled to realize a substantial overlap
of the two condensates, i.e., we employ a deep harmonic trap.
In the practical calculation, we put $U=1.1, \ R=1.0$, and vary the Rabi coupling $g$ to
observe how the condensates change.
The results of the calculation are shown in Fig.\ref{repRabi}.

In Fig.\ref{repRabi},
as a result of the strong magnetic field, there exist vortices in each
condensate.
For the case of weak Rabi coupling as $g=0.01$ and $g=0.02$,
vortex cores of the two condensates are separated.
As $g$ is increased, the locations of cores are getting closer, 
and for $g>0.05$ they coincide.

The above observation is useful to understand properties of the
condensates trapped in two different but finite-overlap potentials,
which are investigated in the following section.


\section{BEC$\mbox{s}$ in harmonic and annular traps:
Rabi coupling and instability of BEC$\mbox{s}$}

In the present section, we shall study dynamics of two BECs trapped by the harmonic
and annular potentials, respectively.
We call the atom trapped the harmonic (annular) potential $A$-atom ($B$-atom)
as before, and these atoms are coherently transferred by the Rabi coupling. 
In the practical calculation, we fixed the inter and intra-species repulsions
$U$ and $R$, and then we varied the Rabi coupling $g$ as a
controllable parameter.
At $t=0$, we suddenly add the Rabi coupling and then we simulate
how the condensates evolve.
We show the results for the case with $f=0.02$, $U=0$ and $R=3.0$, although similar results were obtained in the case of $U>0$.

\begin{figure}[ht]
\begin{center}
\includegraphics[width=4cm]{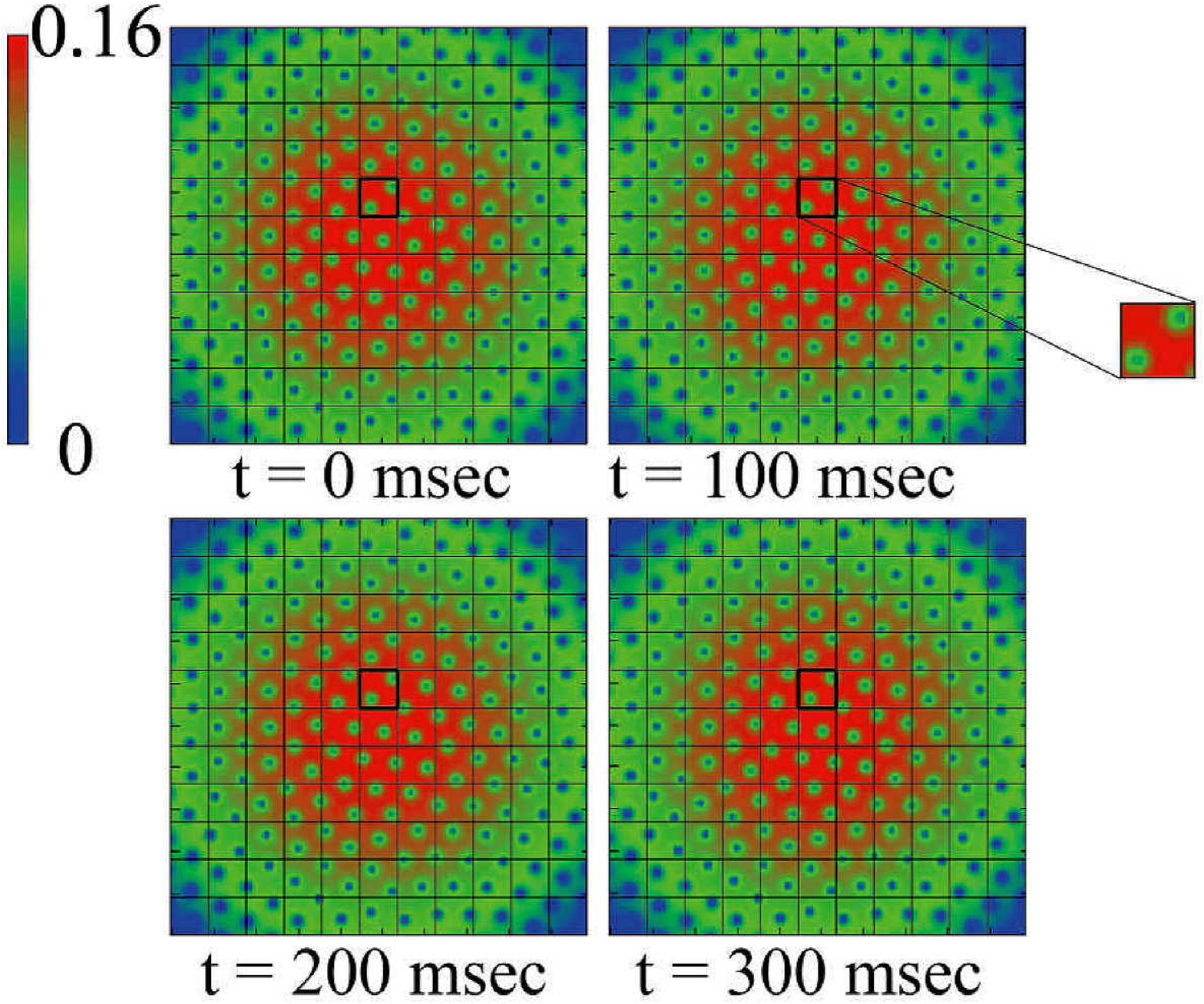} 
\includegraphics[width=4cm]{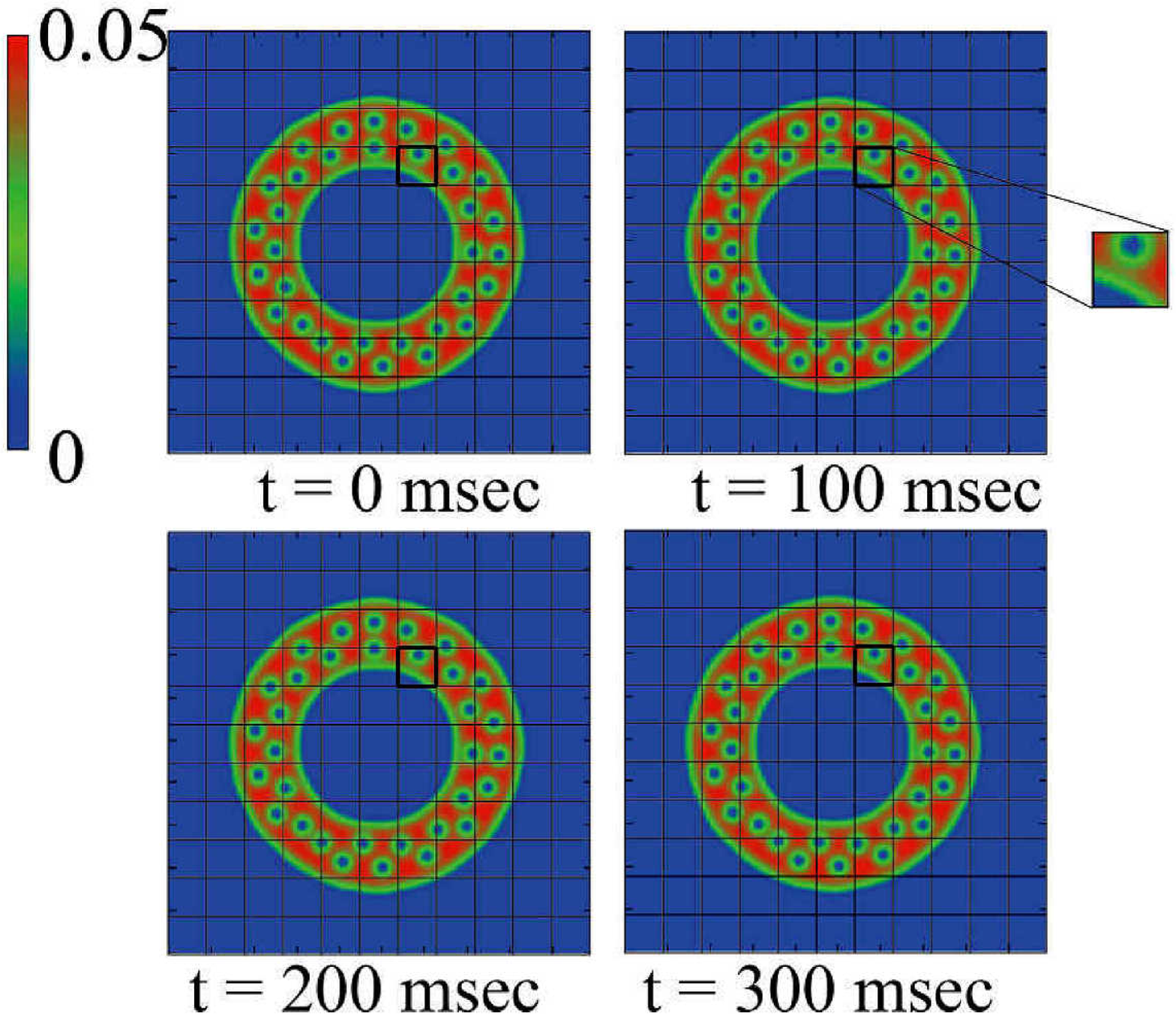} 
\caption{(Color online)
Density profiles of condensates in the harmonic and annular traps as a 
function of time.
Rabi coupling $g=0.01$ is added at $t=0$.
Both BECs reach stable states in which stable vortex solids form
and vortex locations are correlated with each other.
$f=0.02$.
}
\label{ABRabi1}
\end{center} 
\end{figure}
\begin{figure}[ht]
\begin{center}
\includegraphics[width=7cm]{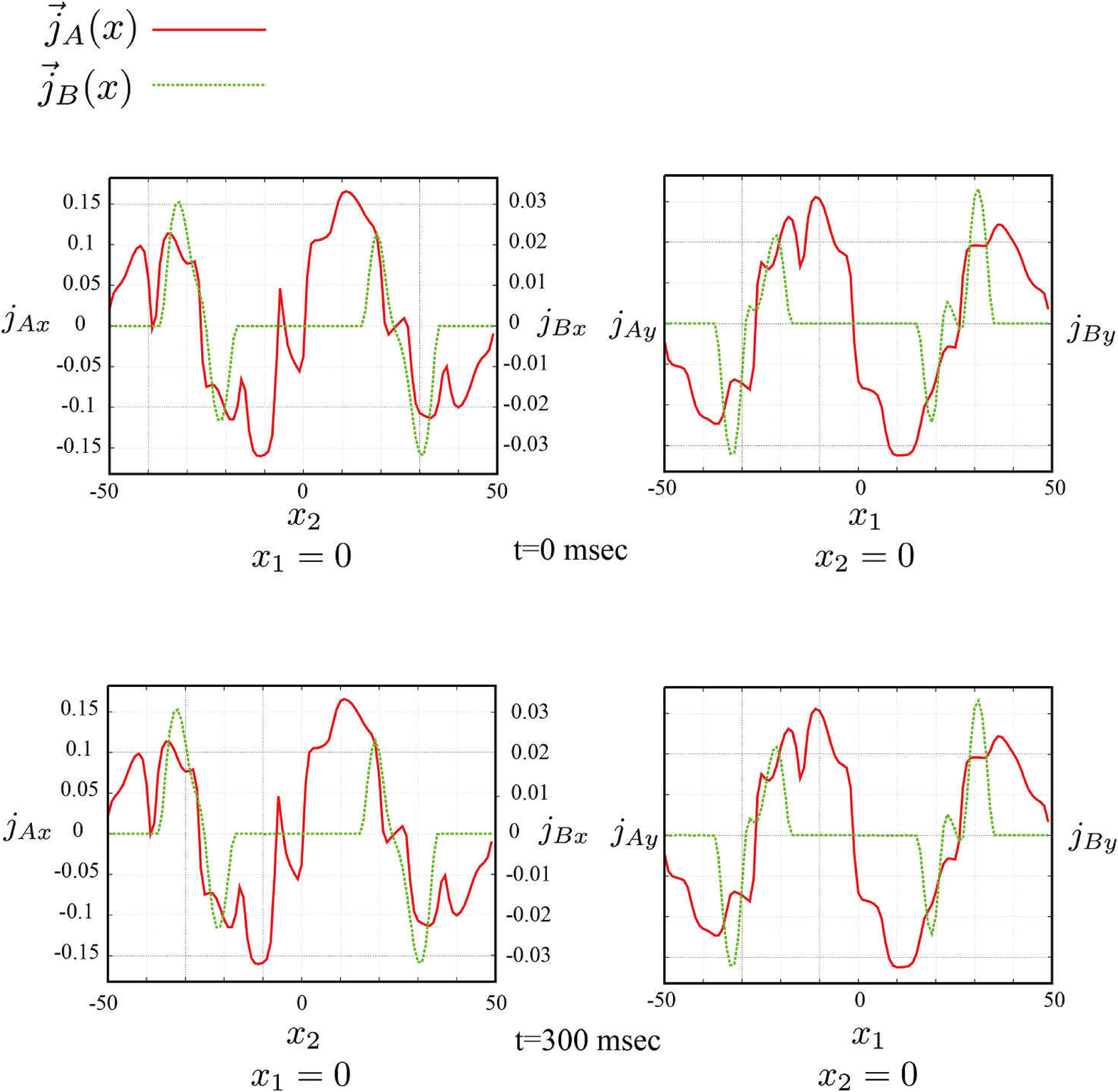} 
\caption{(Color online)
Time evolution of the supercurrents after a sudden turning on the
Rabi coupling $g=0.01$.
The existence of the Rabi coupling does not substantially influence
behavior of the supercurrents.}
\label{SupRabi1}
\end{center} 
\end{figure}

We first show the results for the case of a weak Rabi coupling, i.e.,
$g=0.01$.
See Fig.\ref{ABRabi1} for time evolution of the condensates.
After a sudden turning on the Rabi coupling,
both the BECs form stable states in which vortex solid is stable.
We also show the behavior of the supercurrents in Fig.\ref{SupRabi1}.
It is obvious that the existence of the Rabi coupling does not substantially 
influence behavior of the supercurrents.
The ring-shaped condensate exhibits surface supercurrents near
the inner and outer boundaries, whereas the condensate in the harmonic
trap has no specific pattern of the supercurrent although the supercurrent has
a strong correlation with the vortex locations in the bulk of the
condensate. 

\begin{figure}[ht]
\begin{center}
\includegraphics[width=4.2cm]{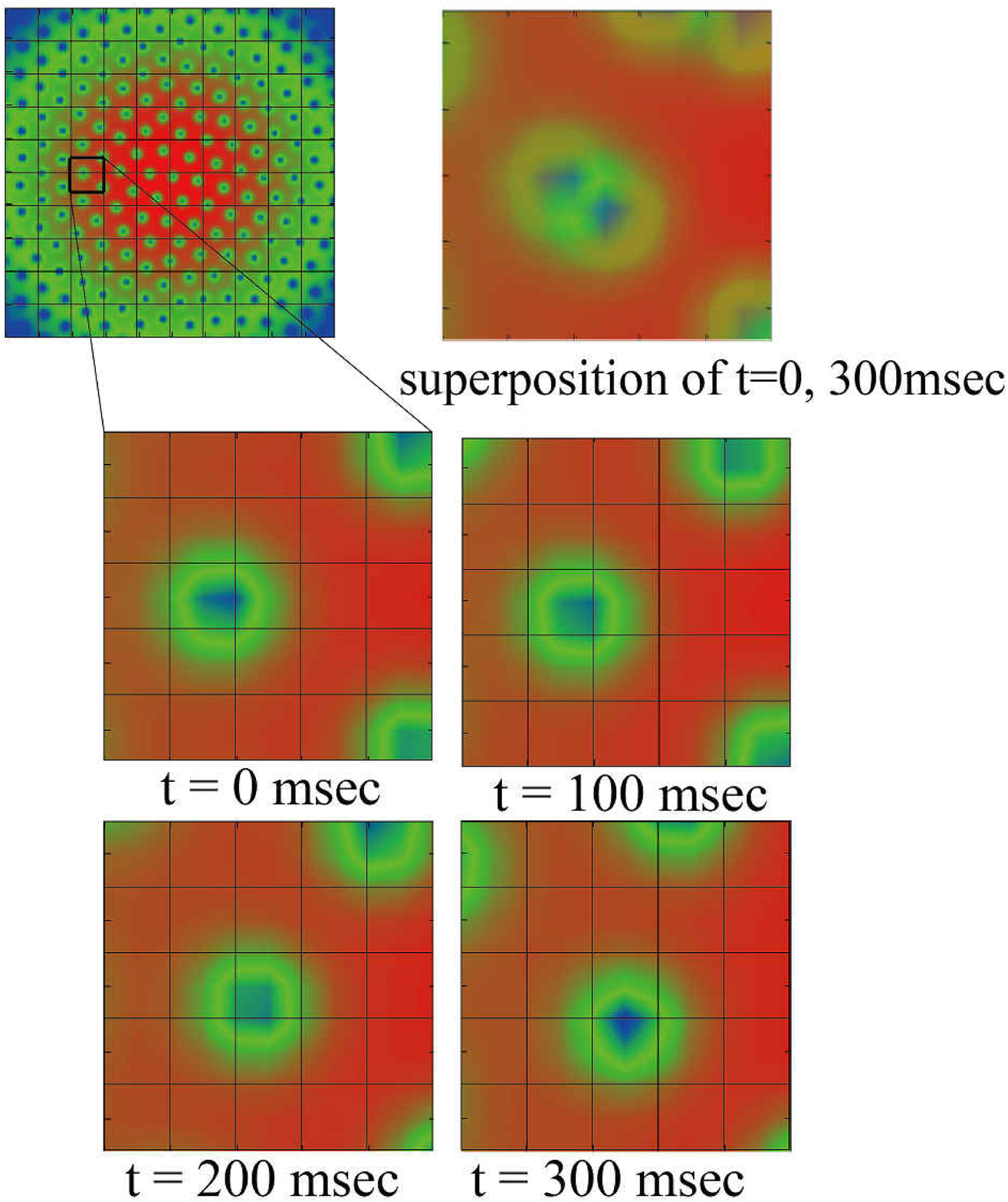} 
\includegraphics[width=4.2cm]{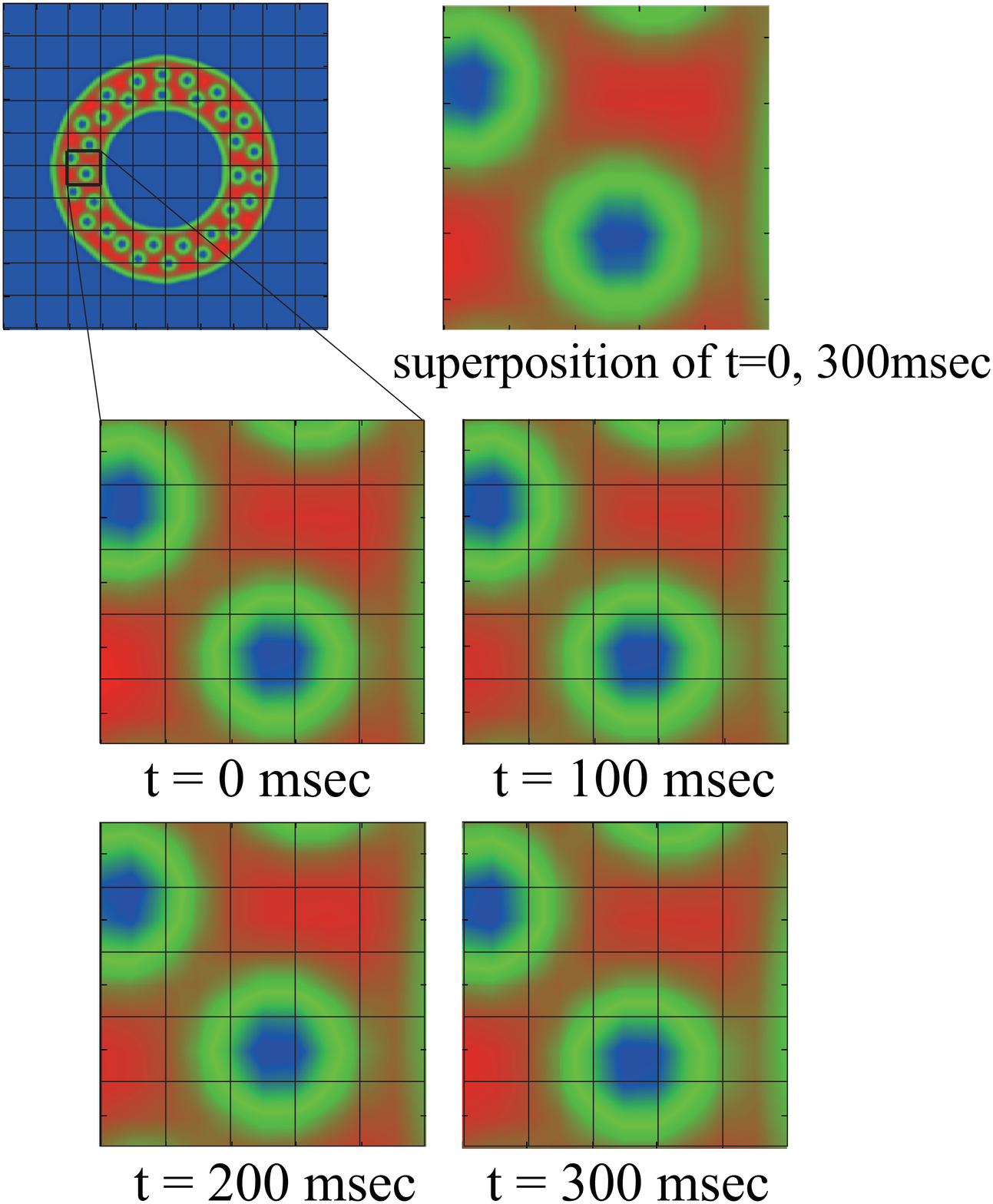} 
\caption{(Color online)
Beahvior of vortices after a sudden turning on the Rabbi coupling $g=0.05$.
Vortices vibrate.
}
\label{vortexRabi1}
\end{center} 
\end{figure}
\begin{figure}[ht]
\begin{center}
\includegraphics[width=7.5cm]{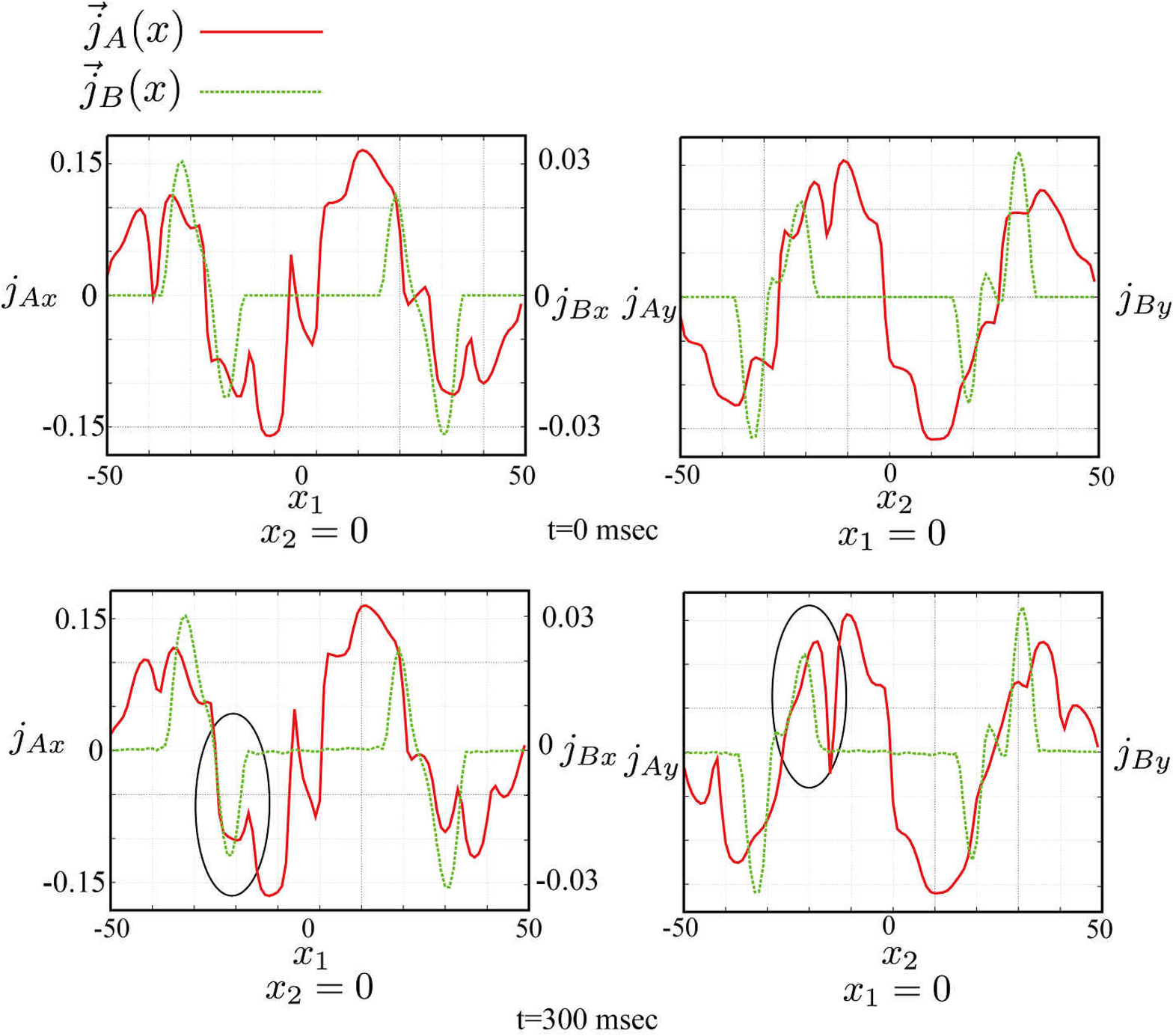} 
\caption{(Color online)
Supercurrents at $t=0$ msec and $300$ msec.
At $t=300$ msec, it is observed that the $A$ and $B$ supercurrents synchronize
with each other.
Rabi coupling $g=0.05$.
}
\label{SupRabi2}
\end{center} 
\end{figure}

\begin{figure}[ht]
\begin{center} 
\includegraphics[width=6.5cm]{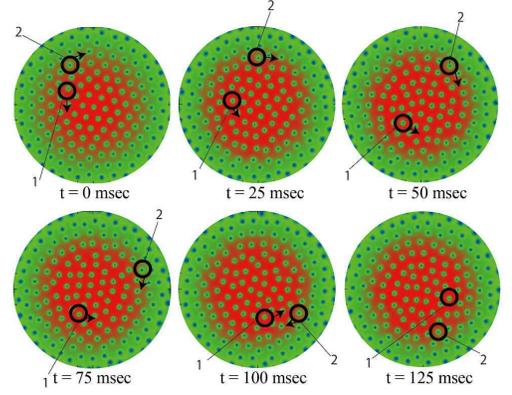} 
\caption{(Color online)
Vortices of $A$-atom condensate move along the boundaries
of the ring-shaped $B$-atom condensate.
In the outer (inner) boundary, vortices move clockwise (counter-clockwise).
Rabi coupling $g=0.1$.
}
\label{vortexRabi2}
\end{center} 
\end{figure}
\begin{figure}[ht]
\begin{center}
\includegraphics[width=7.5cm]{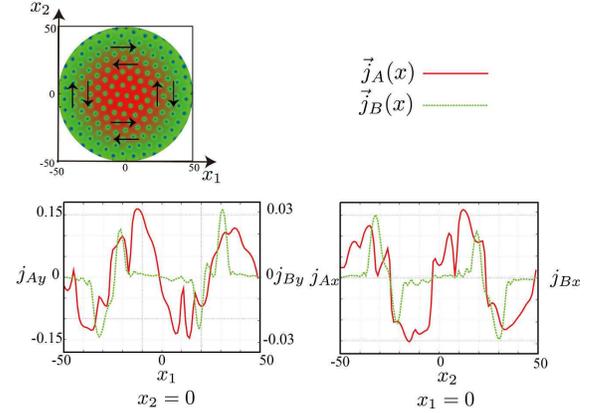} 
\caption{(Color online)
Supercurrent generated in $A$-atom condensate by the Rabi coupling.
Direct movement of $A$-atom vortices induces the clockwise
and counter-clockwise supercurrents that are correlated to the supercurrents 
in the $B$-atom
condensate.
Rabi coupling $g=0.1$.
}   
\label{SupRabi3}
\end{center} 
\end{figure}
\begin{figure}[ht]
\begin{center}
\includegraphics[width=4cm]{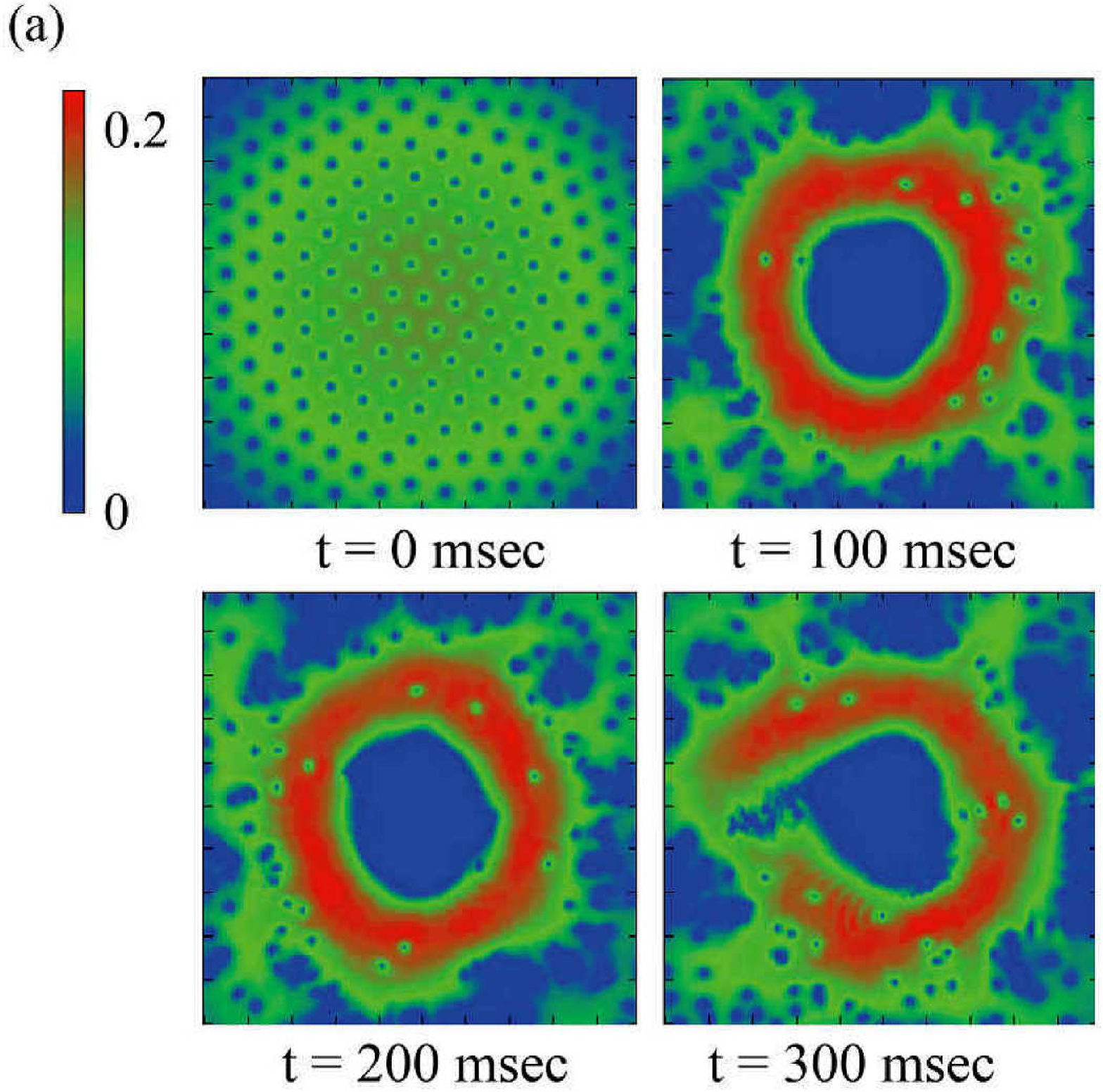} \vspace{0.5cm}
\includegraphics[width=4cm]{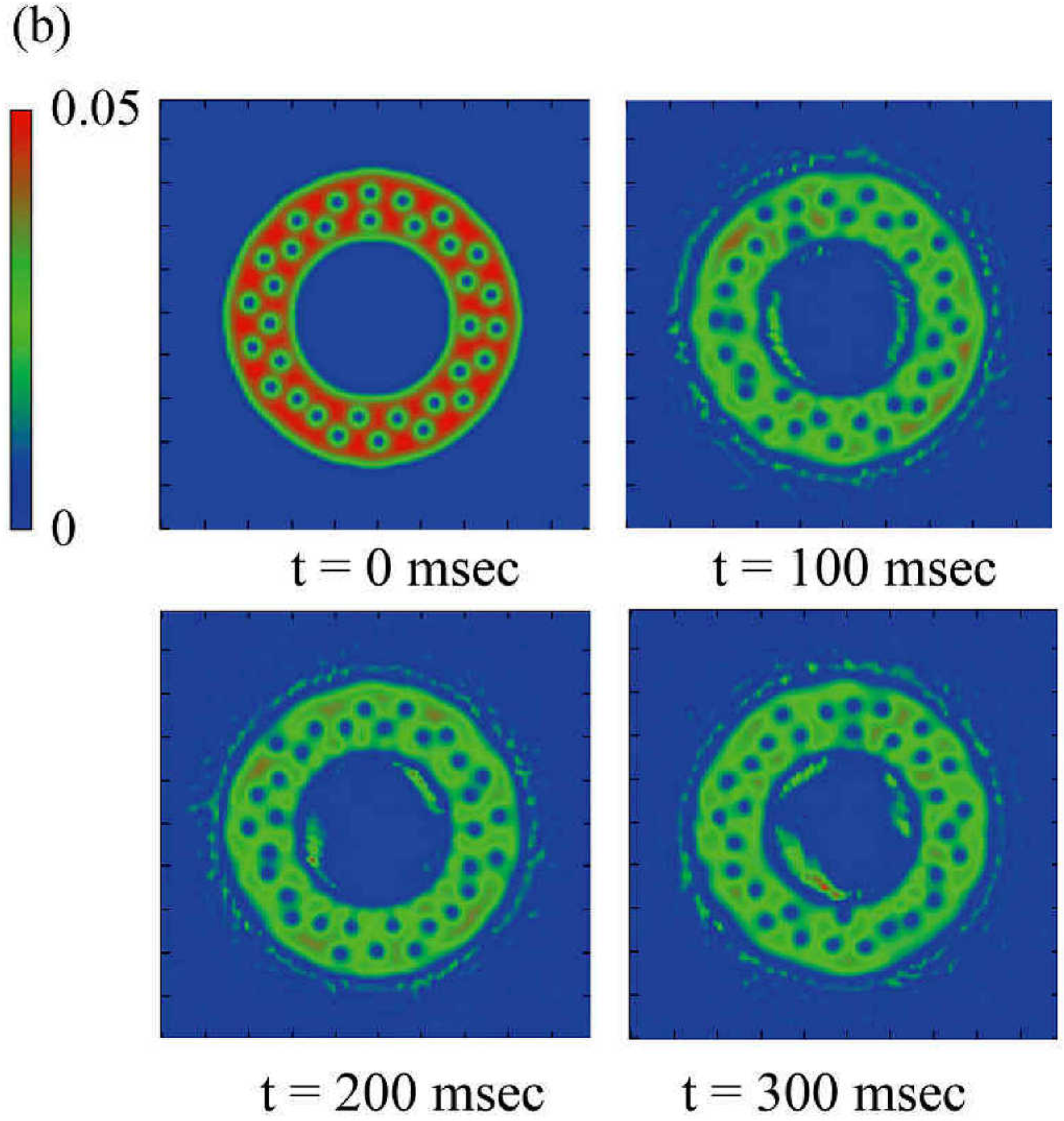} 
\includegraphics[width=8cm]{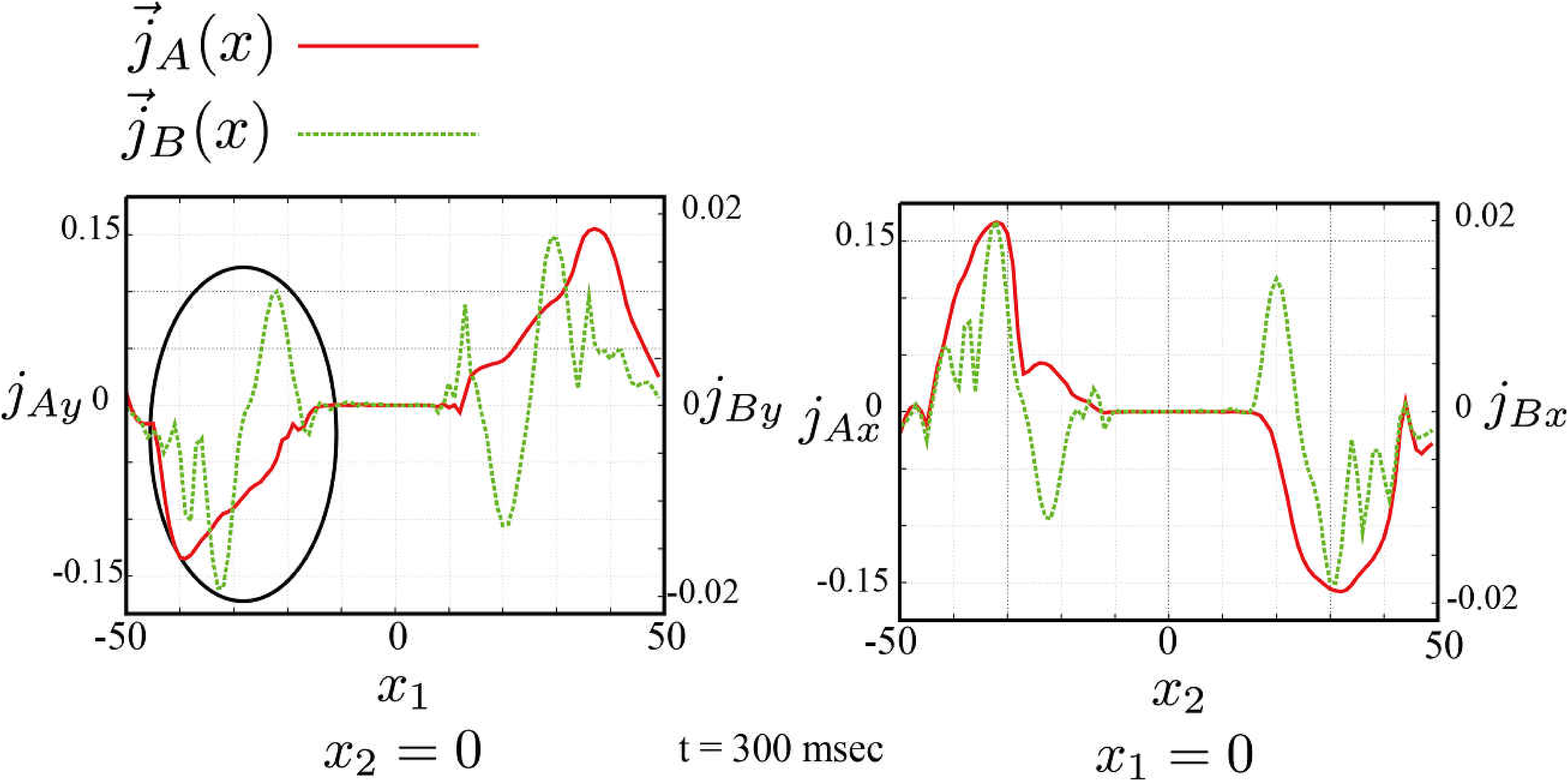} 
\caption{(Color online)
Densities of the condensates and supercurrents in the annular region of
both $A$ and $B$-atom condensates.
Rabi coupling $g=0.25$.
As a result of relatively large Rabbi coupling, $A$ atoms concentrates in the
region of the annular trap.
}
\label{SupRabi4}
\end{center} 
\end{figure}
\begin{figure}[h]
\begin{center}
\includegraphics[width=4cm]{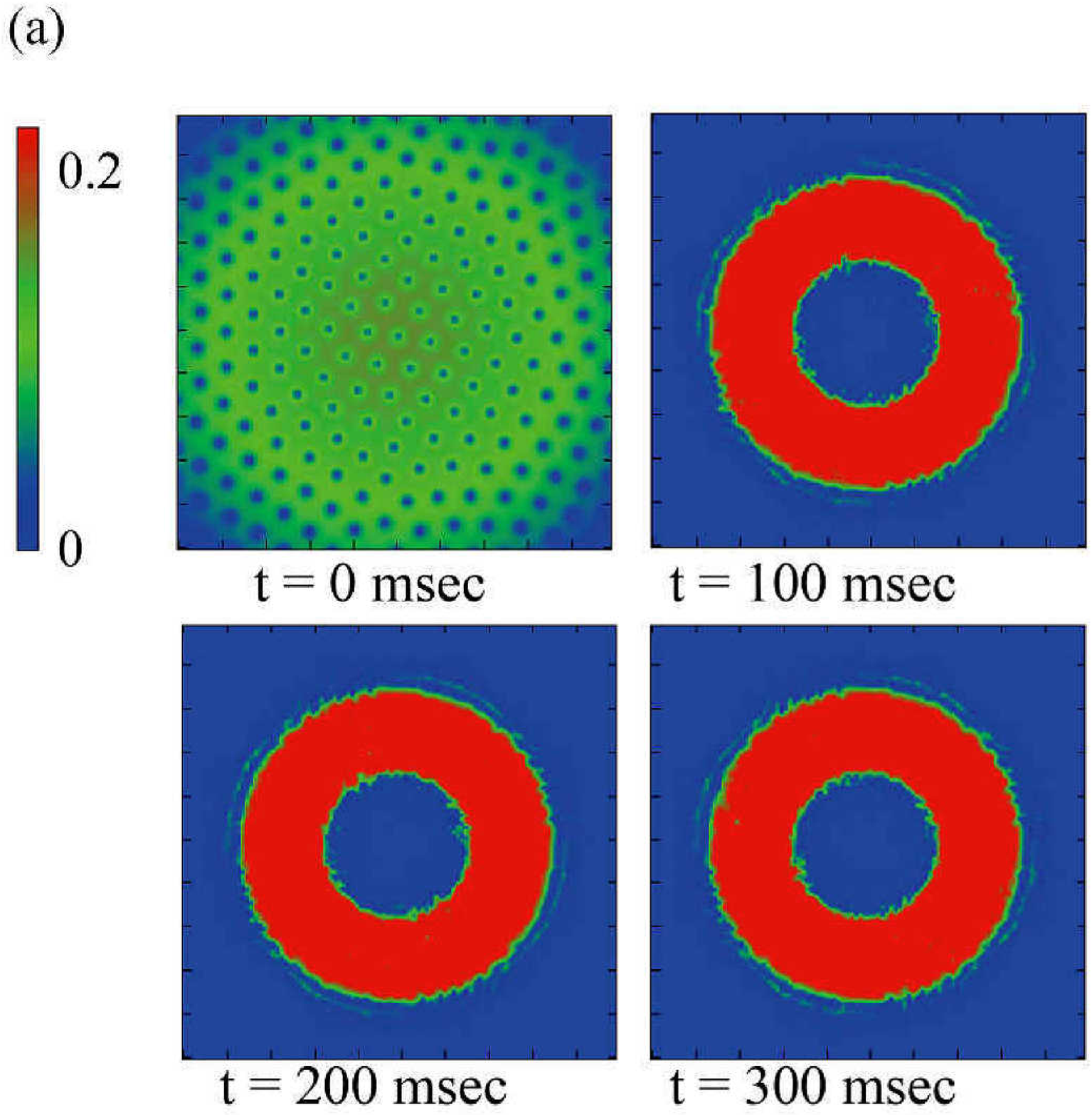} \vspace{0.5cm}
\includegraphics[width=4cm]{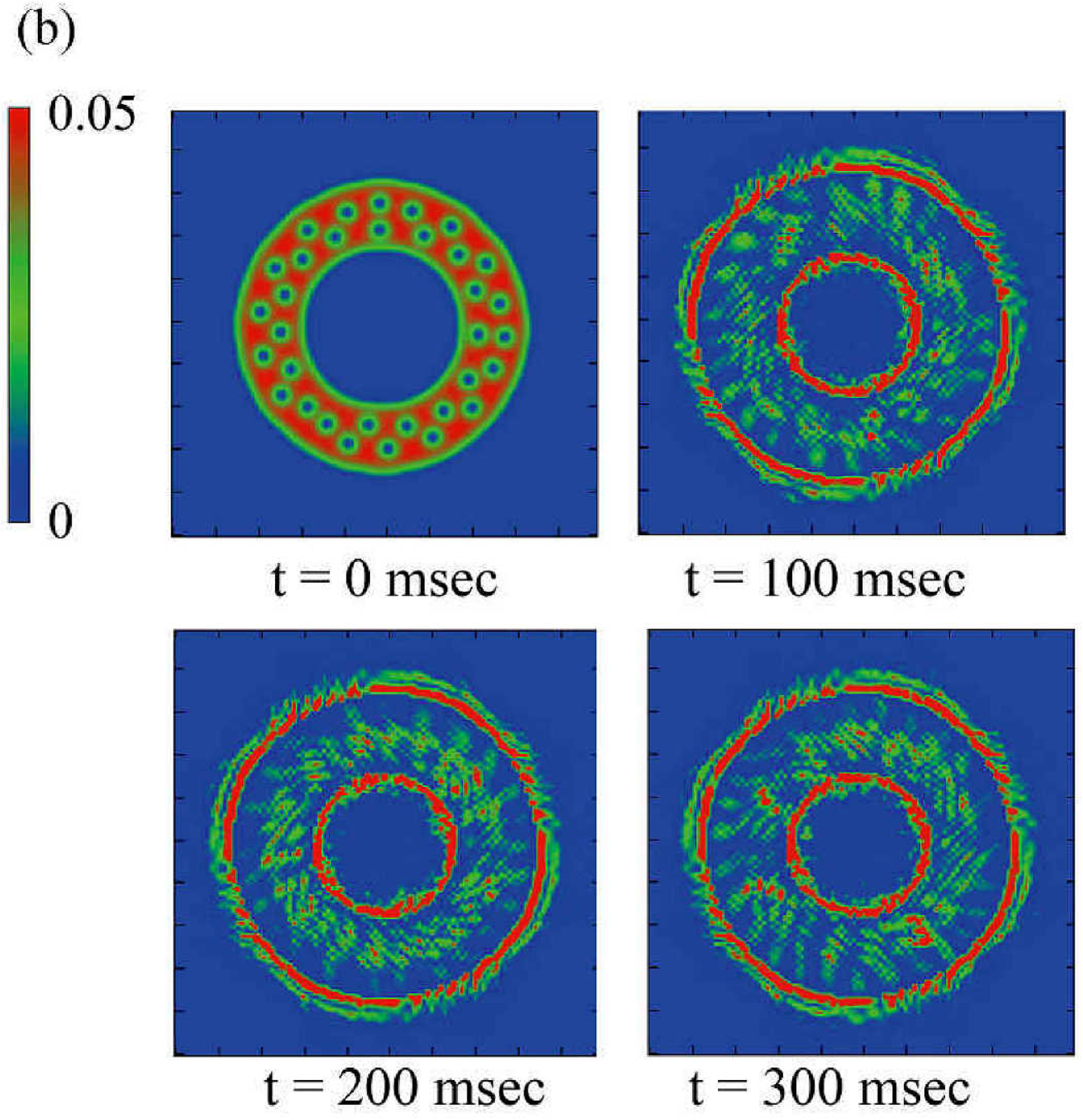} 
\includegraphics[width=8cm]{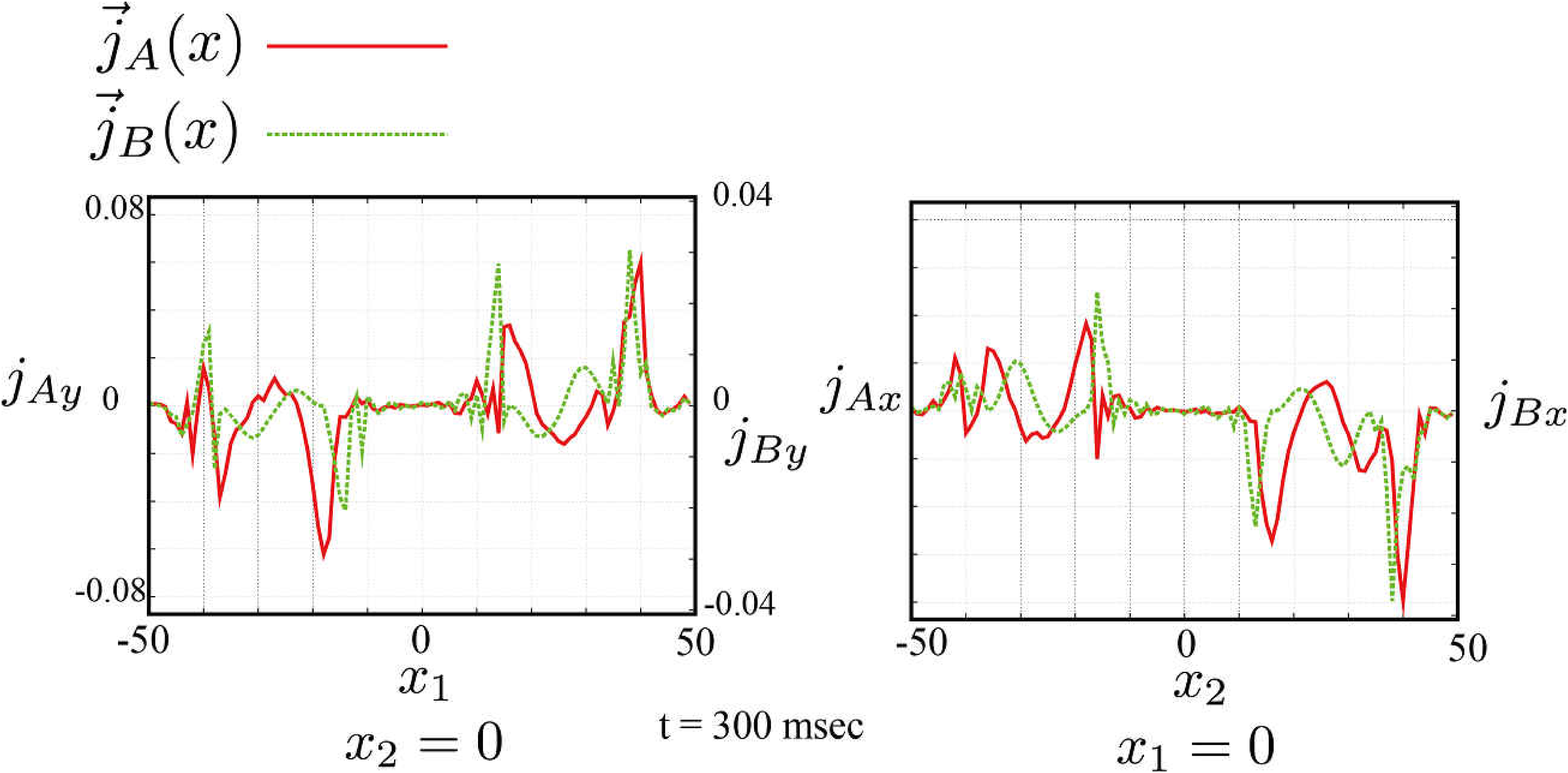} 
\caption{(Color online)
Densities of the condensates and supercurrents in the annular region of
both $A$ and $B$-atom condensates.
Rabbi coupling $g=1.0$.
As a result of large Rabbi coupling, $A$ atoms concentrates in the
region of the annular trap, and $B$ atoms have a substantial density only near the 
boundaries of the annulus.
No vortices are observed and there exit no stable supercurrent.
This result indicates that coherent BECs become unstable by the large
Rabbi coupling.
}
\label{SupRabi5}
\end{center} 
\end{figure}

Next let us consider the case $g=0.05$.
By the practical calculation, we observed time evolution of the 
condensates, in particular, we were interested in behavior of vortices.
See Fig.\ref{vortexRabi1} for the numerical results.
We found that vortices vibrate around their equilibrium locations.
Precise investigation shows that vibration of $A$-atom vortices is
larger than that of $B$-atoms.
By the Rabi coupling, transference of $A$ and $B$-atoms takes place,
and as a result, an effective interaction between $A$-vortex and $B$-vortex
appears.
The repulsion between intra-species vortex and the effective inter-species 
vortex attraction by the Rabi coupling compete and they cause the vortex vibration.
In Fig.\ref{SupRabi2}, we show the supercurrents in the condensates
at $t=0$ and $t=300$ msec.
At $t=300$ msec, it is observed that the $A$ and $B$-atom supercurrents
partly synchronize.

For the case of $g=0.1$, we found that vortices of $A$-atom
move along the boundaries of the ring-shaped $B$-atom condensate.
See Fig.\ref{vortexRabi2}.
In the outer (inner) boundary, $A$-vortices move clockwise (counter-clockwise).
This is a result of the Rabi coupling and the surface supercurrents in the 
$B$-atom condensate.
See also supercurrent in the $A$-atom condensate generated by the Rabi
coupling in Fig.\ref{SupRabi3}.
It can be seen that the supercurrents in the $B$-atom condensate
induce the ring-shaped supercurrents in the $A$-atom condensate.
Direct movement of $A$-atom vortices generates the clockwise
and counter-clockwise superflows correlated to those in the $B$-atom
condensate.

We furthermore increased the Rabi coupling $g$ to $g=0.25$.
Densities of the condensates and supercurrents are shown in Fig.\ref{SupRabi4}.
As a result of relatively large Rabi coupling, $A$-atom tends to concentrate in the
region of the annular trap, and the density of $A$-atom varies as a function
of time.
On the other hand, the $B$-atom condensate seems relatively stable.
It is very interesting to notice that the $A$-atom condensate has a supercurrent
in the ring azimuthal direction and it flows only in one direction, i.e., clockwise,
whereas the $B$-atom condensate has surface supercurrents flowing clockwise
and anticlockwise, respectively.

Finally, we show the results for the strong Rabi coupling $g=1$.
See Fig.\ref{SupRabi5}.
$A$-atoms concentrate in the region of the annular trap, and 
$B$-atoms have a large density near the boundaries of the annulus.
Vortices disappear and no stable supercurrents exist.
This result means that coherent BECs become unstable by the large
Rabi coupling.


\section{Conclusion}

In this paper, we studied the two-component Bose gas trapped by
the harmonic and annular potentials.
They interacts with each other by the coherent Rabi coupling and
the inter-repulsions.
The GPEs were used for studying low-energy states of the system.

We first investigated single-component bosons in the harmonic and annular
traps, respectively.
In particular, the detailed study on the ring-shaped condensate in the annular
trap showed that vortices exhibit interesting behavior by applying an
artificial magnetic field.
We also studied the effect of the WL located in the annular condensate,
and obtained the phase of the BEC by solving GPE.
Compared with the previous calculation in the 1D case, we found the unexpected
behavior of the phase of the condensate with the WL. 
We measured the supercurrents and showed that they exhibit various different
behaviors depending on the width of the annular condensate.

Finally, we investigated behavior of the BECs in the coupled system with the Rabi coupling
and repulsions between the $A$ and $B$-atoms.
As the strength of the Rabi coupling is increased, the Bose gas system
exhibits various ``phase transition".
First $A$ and $B$-atom vortices start to correlate and they vibrate around
their equilibrium locations of the vortex lattice.
As the Rabi coupling is increased, vortices of the $A$-atom move along the
boundary of the ring-shaped $B$-condensate.
This phenomenon occurs as a result of the superflow at the boundaries
of the $B$-atom BEC and the Rabi coupling.
This observation indicates an interesting possible method to control vortex
movement in the condensate.
We found that for a very large Rabi coupling BECs become unstable
as the vortex flow disturbs the $A$-atom BEC and then it induces 
instability of the $B$-atom BEC again via the large Rabi coupling.

The results in the present study have been obtained by solving the GPEs.
It is desirable to see how the fluctuations, which are not taken into account
in the GPEs, change the results.
The path-integral Monte-Carlo (MC) simulations are useful for this study,
although the time evolution of the system cannot be investigated by the MC simulations.
The MC simulations on the present system are under study, and we hope that
the results will be reported in the near future.

\acknowledgments 
This work was partially supported by Grant-in-Aid
for Scientific Research from Japan Society for the 
Promotion of Science under Grant No.26400246.


\end{document}